\documentclass[structabstract]{aa}

\usepackage{graphicx}
\usepackage{txfonts}
 \usepackage{longtable}
\usepackage{natbib}
\begin{document}
   \title{Variable gamma-ray sky at 1 GeV}


   \author{M.S. Pshirkov\thanks{pshirkov@ulb.ac.be} \inst{1,3}
\and
      G.I. Rubtsov \thanks{grisha@ms2.inr.ac.ru}\inst{2}
          }

   \institute{Universite Libre de Bruxelles, Service de Physique Theorique, CP225, 1050,  Brussels, Belgium
    \and
    Institute for Nuclear Research of the Russian Academy of Sciences, 117312, Moscow, Russia
         \and
             Pushchino Radio Astronomy Observatory of Lebedev Physical Institute, 142290, Pushchino,
Russia              }

   \date{}


   \abstract
{}
  {We search for the long-term variability of the $\gamma$-ray sky
in the energy range $E >$ 1 GeV with 168 weeks of Fermi-LAT data. }
  {We perform a full sky blind search for regions with variable flux
    looking for deviations from uniformity. We bin the sky into
    12288 bins using Healpix package and use Kolmogorov-Smirnov test
    to compare weekly photon counts in each bin with a constant flux
    hypothesis. The weekly exposure of Fermi-LAT for each bin is
    calculated with the Fermi-LAT tools. We consider flux variations in
    the bin significant if statistical probability of uniformity is
    less than $4\times 10^{-6}$, which corresponds to 0.05 false
    detections in the whole set.}
{ We identified 117 variable sources,
  variability of 27 of which has not been reported before. Among the
  sources with previously unidentified variability there are 25 AGNs
  belonging to blazar class (11 BL Lacs and 14 FSRQs), one AGN of
  uncertain type and one pulsar PSR~J0633+1746~(Geminga). The observed
    long term flux variability of Geminga  has a statistical
    significance of 5.1$\sigma$.}
 {}


   \maketitle
%

\section{Introduction}
\label{sec:intro} Time domain astronomy at different wavelengths
from radio to very high-energy gamma-rays is developing very rapidly
nowadays. In the high energy (HE) range ($\ge 100$ MeV) great progress
was achieved with advent of the gamma-ray telescope Fermi-LAT
\citep{Atwood2009}. Its high sensitivity and almost uniform sky
coverage allow one to study variability of large number of sources at
these energies on time scales from seconds to years.  Sources that demonstrate the most variable behaviour are
active galactic nuclei (AGN), primarily of the blazar type
\citep{Abdo2010a}. It is well known that these sources exhibit
variability on different time scales and at different wavelengths
\citep{Carini1991, Ulrich1997,Welsh2011, Rani2009, Bonning2009,
  Soldi2008, Ghisellini2008, Raiteri2005,
  Ciprini2003,Urry1993}. Studies of time variability of these sources
are very important for better understanding of AGN engines;
they are also essential for assessing the quality of spectral energy
distributions obtained from multiwavelength observations made at
different epochs.

In this paper we perform a full sky blind search for variable
sources. We bin the sky into equal area pixels and search for
deviations of photon number counts from the uniformity in time.


\section{Data and method}
\label{sec:data_method}


%
The LAT Pass 7 weekly all-sky data publicly
available at Fermi mission
website\footnote{http://fermi.gsfc.nasa.gov/ssc/data/access/} were
used in this work. The
analysis covers the time period of 168 weeks from August 04, 2008
to October 18, 2011, corresponding to mission elapsed time (MET)
from 239557417~s to 340622181~s. We use the 'Pass 7 Source'
event class photons with $E>1$~GeV and impose an Earth relative
zenith angle cut of $100^\circ$ and rocking angle cut of
$52^\circ$.

We bin the data week by week using \textsc{HEALPIX} package
\citep{Healpix} into a map of resolution $N_{\mathrm{side}} = 32$ in
galactic coordinates with 'RING' pixel ordering.  Total number of
pixels is equal to 12288 and the area of each pixel is 3.6 sq. deg,
chosen according to the size of Fermi-LAT point-spread function (PSF)
above 1 GeV  which is approximately $1^{\circ}$. We estimate integral
weekly exposure for each pixel using the standard Fermi-LAT tools {\it
  gtltcube} and {\it gtexpcube} (ScienceTools-v9r23p1-fssc-20110726).

\begin{figure}
\begin{center}
\includegraphics[width=6.5 cm]{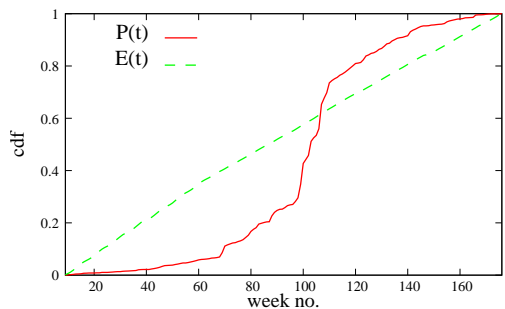}
\end{center}
\caption{Plot of the  cumulative functions $\mathcal{E}(t)$ and $\mathcal{P}(t)$
for pixel no.54 ($l=261^{\circ}, ~b=82.^{\circ}69$). The
difference could be easily seen and probability that the photon
rate is constant is only $P=4\times10^{-80}$. }
 \label{fig:cdf}
\end{figure}

\begin{figure}
\begin{center}
\includegraphics[width=6.5 cm]{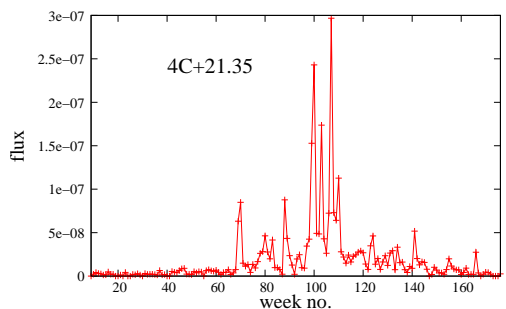}
\end{center}
\caption{Flux of photons with energies larger than 1 GeV for pixel no.54 (4C+21.35, see Figure \ref{fig:cdf}). The flux is in $\mathrm{photons~cm^{-2}~s^{-1}}$ units. }  \label{fig:flux}
\end{figure}

 For each pixel we count the number of photons in each of 168 weeks and
consider corresponding values of weekly exposure. Typically there are 2--10 photons in a pixel per week except bins with the brightest sources. Cumulative
distribution functions (CDFs) $\mathcal{P}(t),\mathcal{E}(t)$ for both
photon counts and exposure are constructed. In the absence of
variability, the photon counts would represent a random process with CDF
proportional to $\mathcal{E}(t)$ and thus $\mathcal{P}(t)$ would
follow $\mathcal{E}(t)$ with deviations caused by a finite number of
observed photons. Otherwise $\mathcal{P}(t)$ would not be
statistically compatible with $\mathcal{E}(t)$. A Kolmogorov-Smirnov
(KS) test is a natural and straightforward way to examine statistical
compatibility of the observed photon counts with the distribution
given by CDF $\mathcal{E}(t)$. The probability that both sets
represent the same distribution could be estimated from the maximal
value of distance between the functions $\mathcal{P}(t)$ and
$\mathcal{E}(t)$.  An example CDFs for one of the pixels is shown
  in Figure~\ref{fig:cdf} and the corresponding flux is shown in
  Figure~\ref{fig:flux}.

 Implemented KS test  is most sensitive to the variability at  long scales (longer than a week),
while transient bursts and flares at shorter time scales may be missed
if they are not overwhelmingly strong. On the other hand, our method
is sensitive to gradual moderate changes in photon fluxes without any
prominent bursts, which could be missed by burst searching techniques.

\begin{figure}
\begin{center}
\includegraphics[width=17cm]{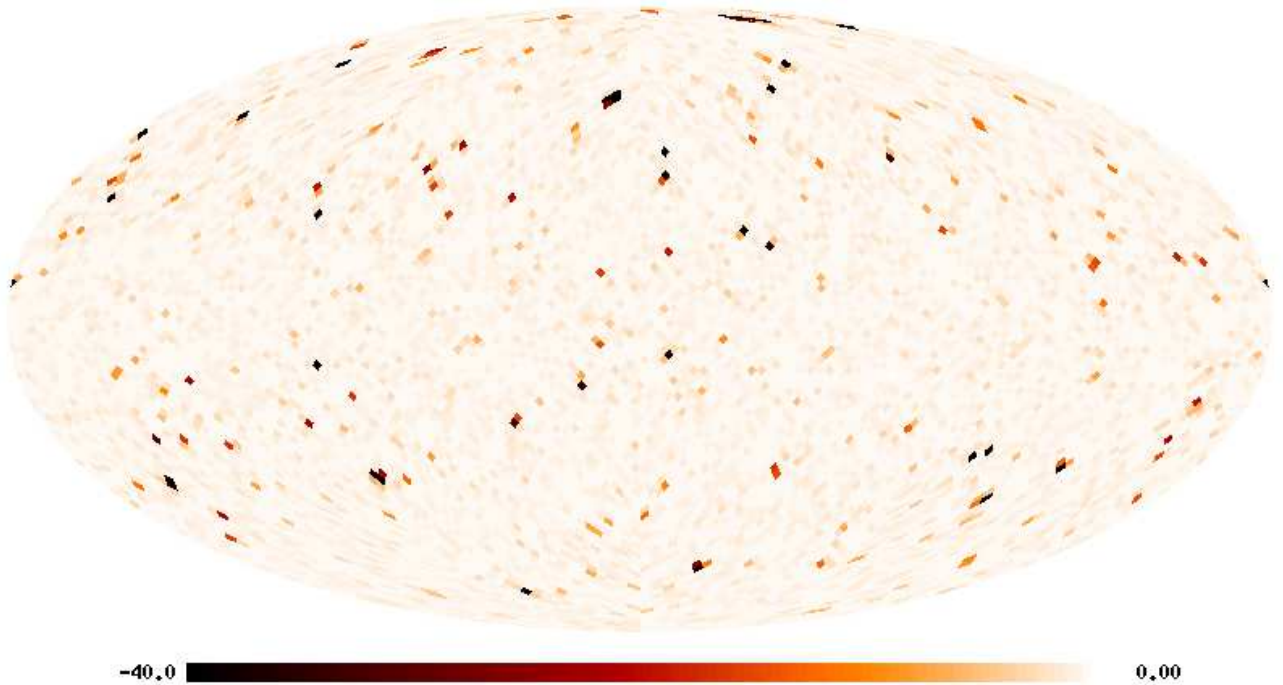}
\end{center}
\caption{Map of Fermi-LAT variability at 1 GeV in galactic
  coordinates. Pixel color represents base 10 logarithm of
  Kolmogorov-Smirnov probability of the uniformity of the observed
  flux. The galactic center is in the center of the figure,
  $l=180^{\circ}$ is on the left.} \label{fig:map}
\end{figure}

KS probability is calculated for each pixel -- we are interested in
pixels with probabilities smaller than the threshold value
$P_0=4\times10^{-6}$. This threshold value is set to allow for penalty
coming from the large number ($N_{\mathrm{pix}}=12288$) of trials: the
detection criterion is chosen in such a way that the entire search
would give a single false detection  with the probability of
  $P_0\times N_{\mathrm{pix}}~\sim~0.05$. A map of the probabilities
is presented in Figure \ref{fig:map}.

\section{Results}
\label{sec:results}
 The total number of bins with probabilities smaller than the threshold
 value $P_0$ is 151. The source identification for each of these bins
 is performed as follows. We consider the photons that arrived during
 several weeks at the epoch of the maximal flux. The center of mass of
 the spatial distribution of these photons is used as an initial
 estimate of the source location. We search for sources from
 2FGL~\citep{2FGL} catalog in the circle with a radius equal to the error
 of center of mass estimation which is usually about 10'-15' for $\sim
 20$ photons). For 34 pixels no source is found with the initial
 estimate: in 31 of them the variability time pattern and photon
 spatial distribution leads us to identification of the source located
 in the neighbouring bin and already identified there. In three cases (pixels 2877 and 2749,
 PMN~J1532-1315, and pixel 2299, SBS~0846+513) we have found no
 counterpart in 2FGL catalog and use Simbad astronomical
 database. Gamma-ray flares from the latter two sources
   (PMN~J1532-1315 and SBS~0846+513) have been recently reported by
   Fermi-LAT~\citep{AT3579,AT3452}, but the activity period is outside
   of the time region of 2FGL catalog.   In order to avoid false
 identifications we performed an additional check in four cases when
 two different sources are residing in adjacent pixels: significant
 difference in the observed luminosity curves confirms that these
 detections are real. The results are presented in Table
 \ref{table:high}.

\begin{figure}
\begin{center}
\includegraphics[width=6.0cm]{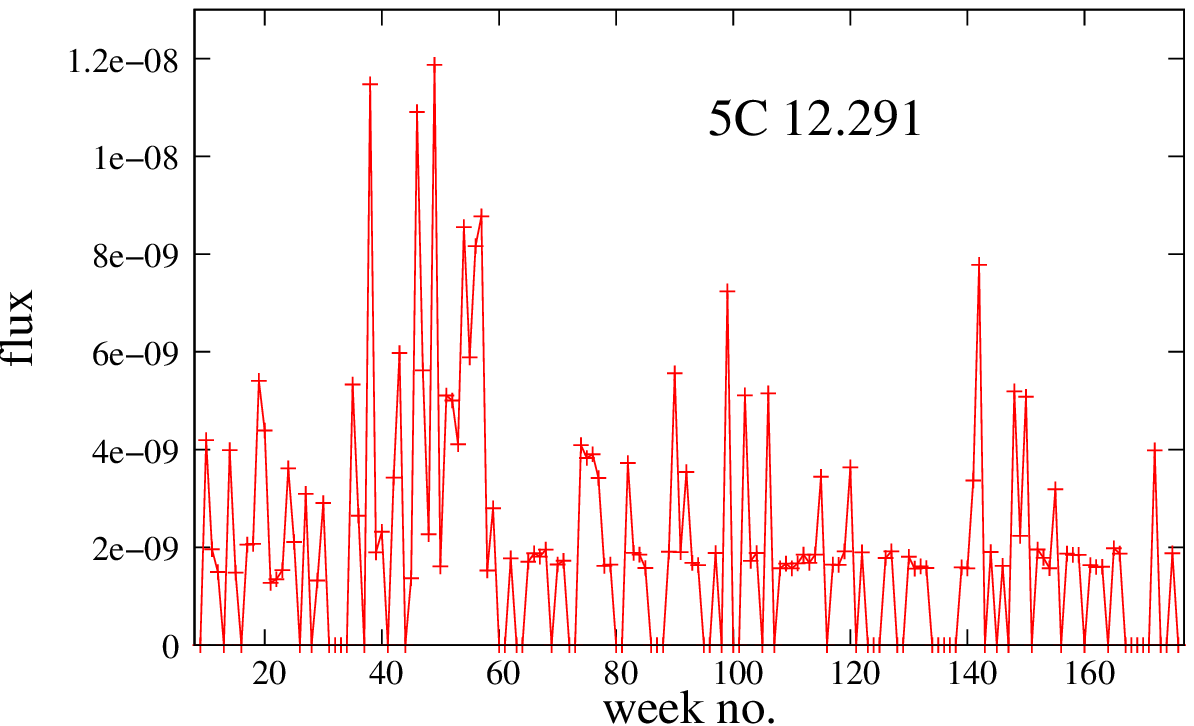}
\includegraphics[width=6.0cm]{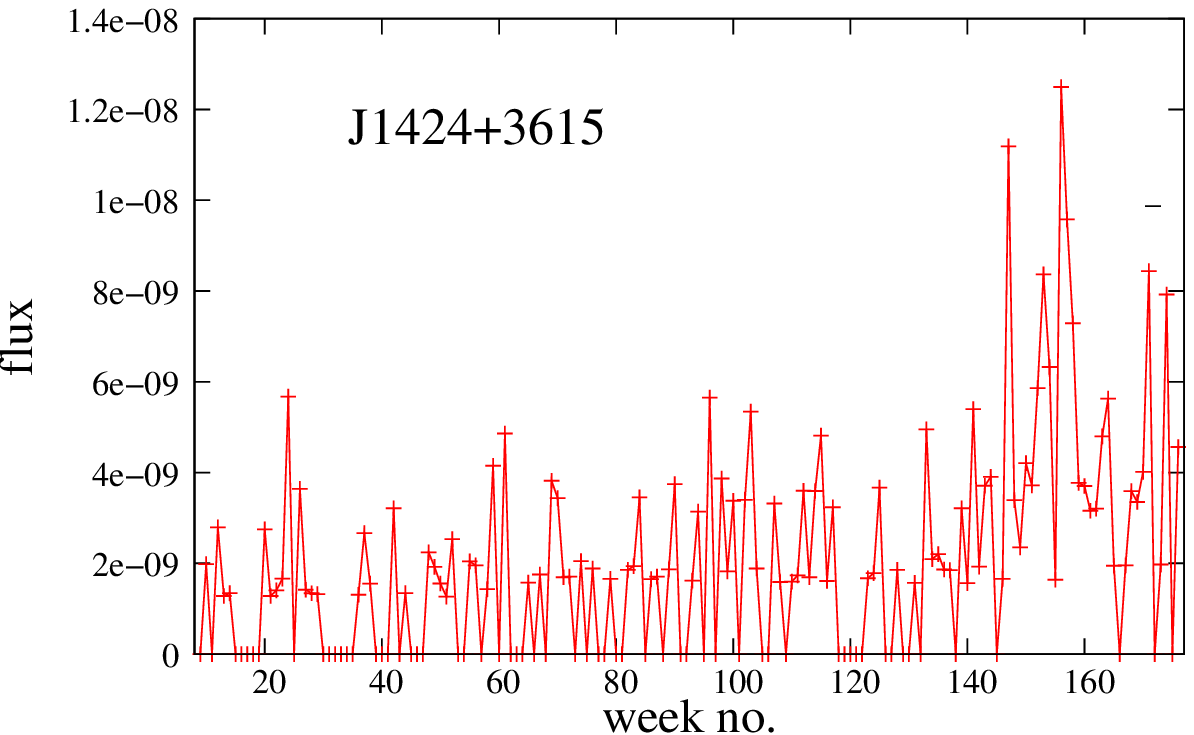}\\
\includegraphics[width=6.0cm]{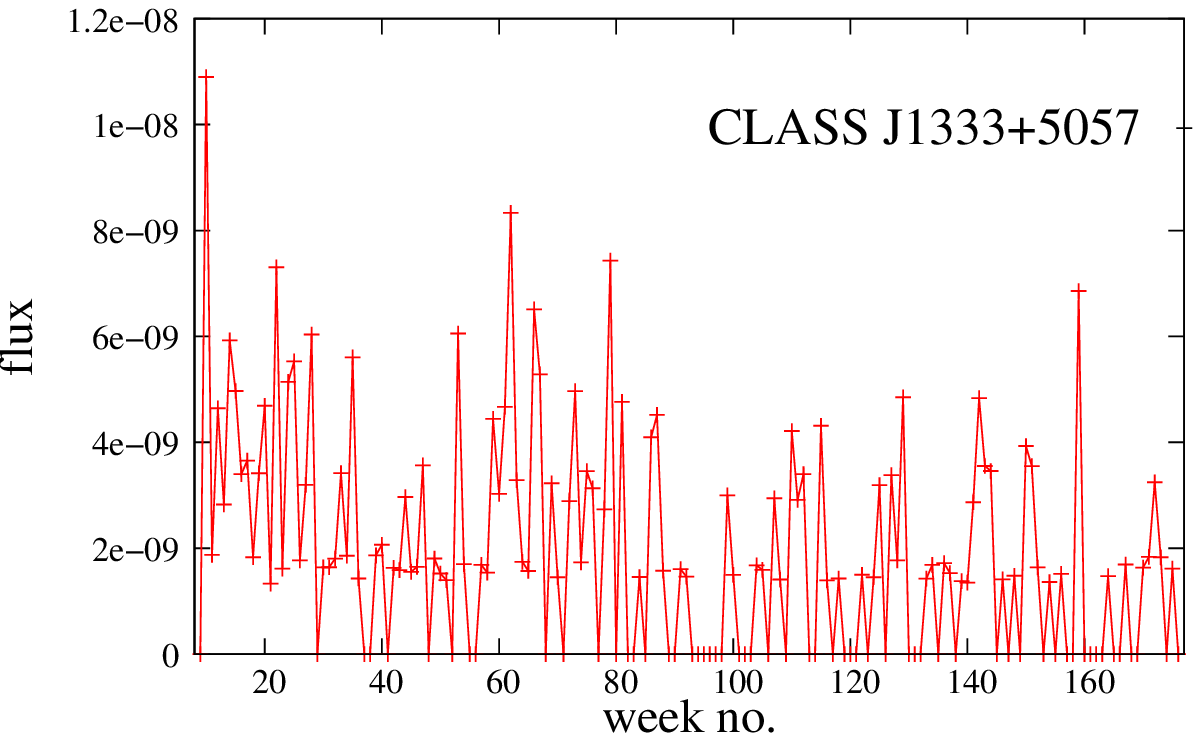}
\includegraphics[width=6.0cm]{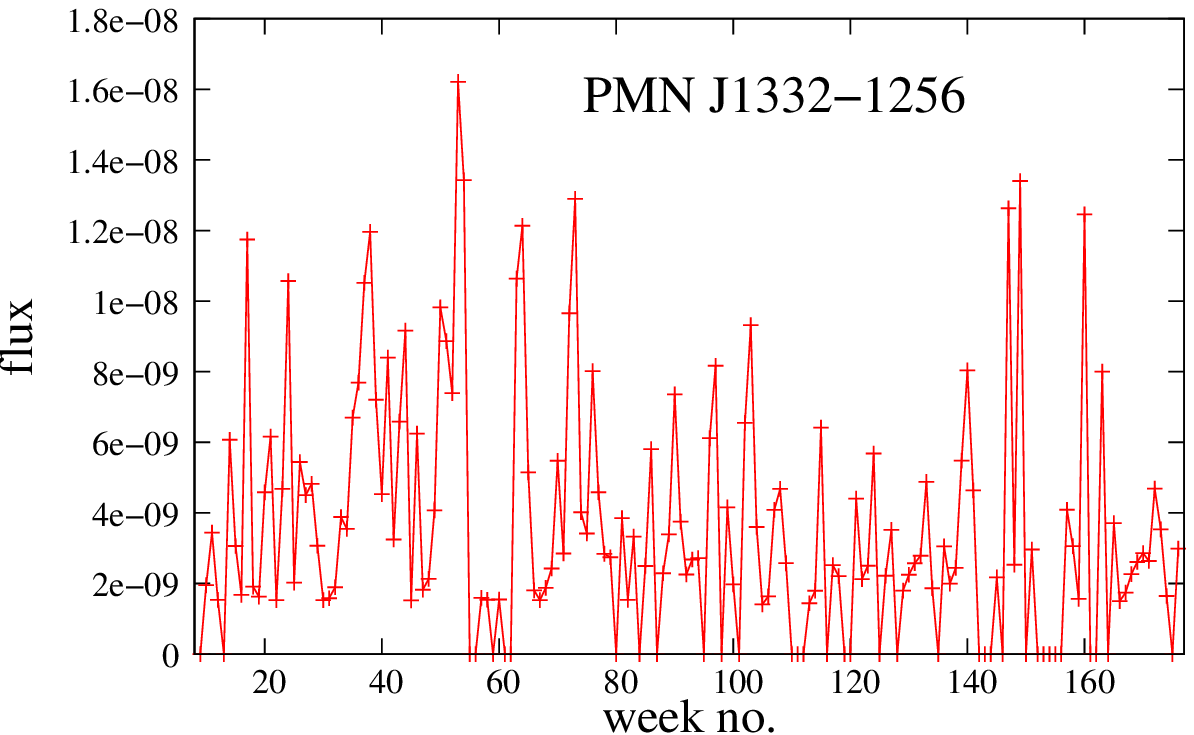}\\
\includegraphics[width=6.0cm]{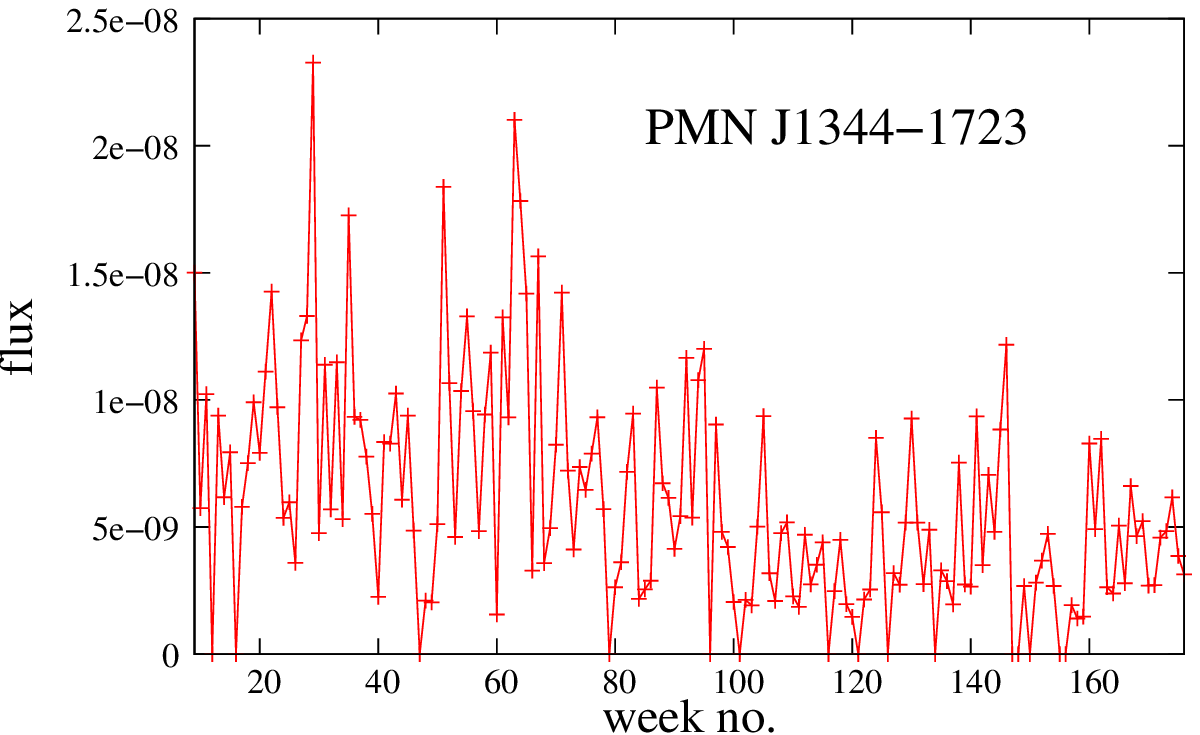}
\includegraphics[width=6.0cm]{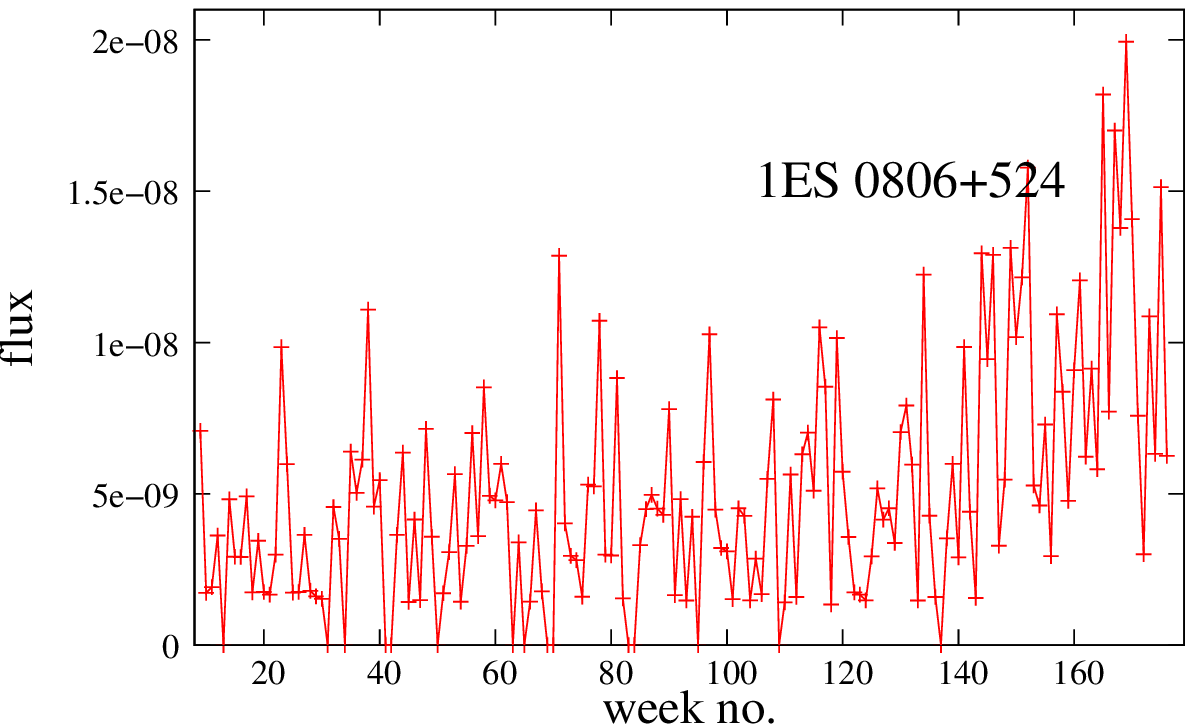}
\end{center}
\caption{Luminosity curves for variable sources listed in the Table \ref{table:undetected}. The flux is in $\mathrm{photons~cm^{-2}~s^{-1}}$ units.  } \label{fig:flux1}
\end{figure}

\begin{figure}
\begin{center}
\includegraphics[width=6.0cm]{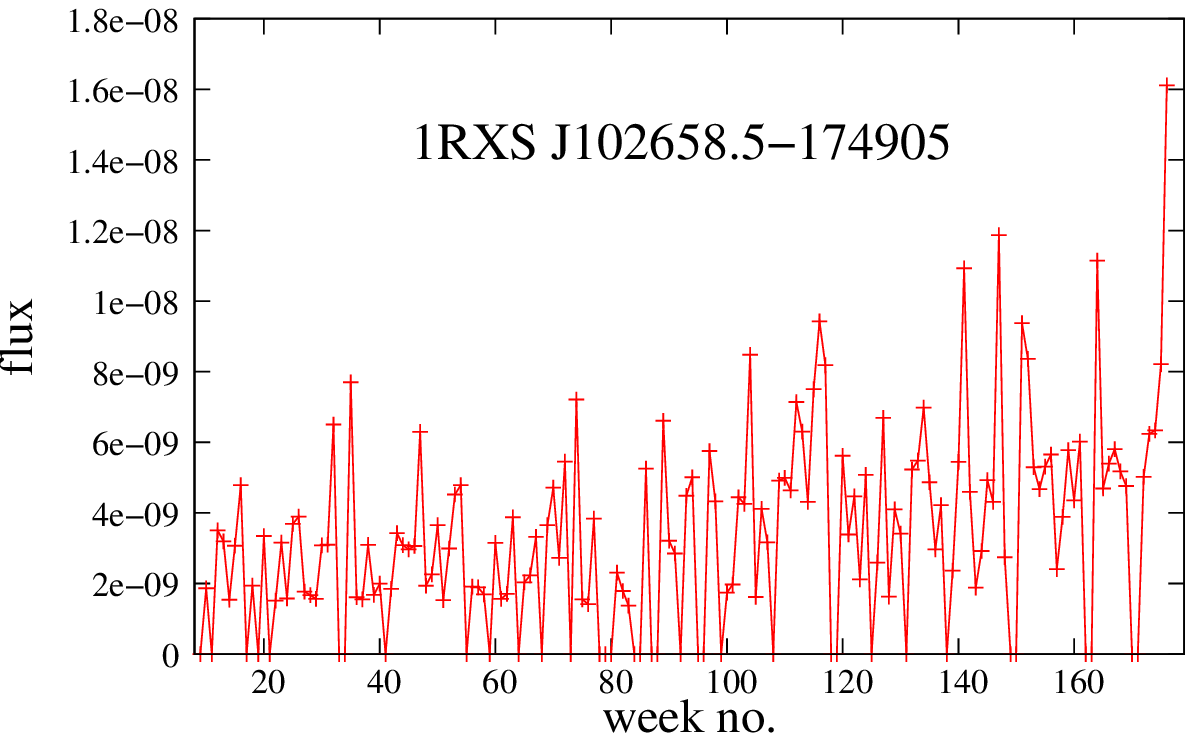}
\includegraphics[width=6.0cm]{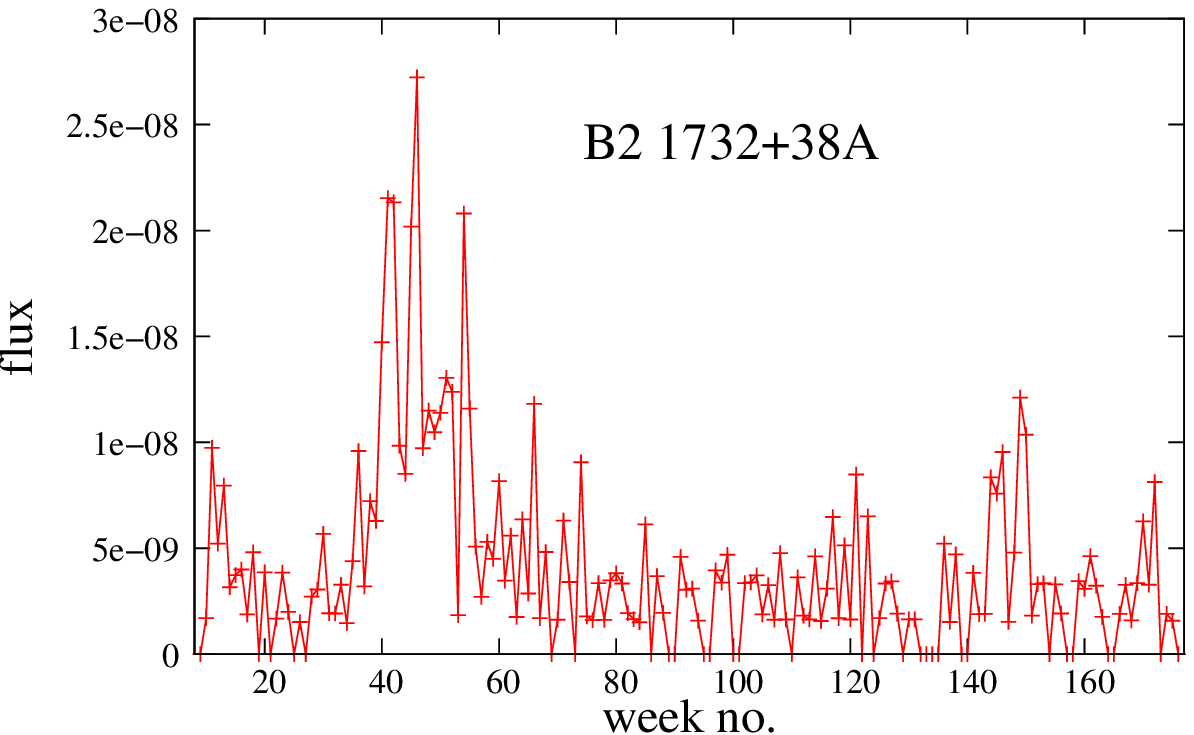}\\
\includegraphics[width=6.0cm]{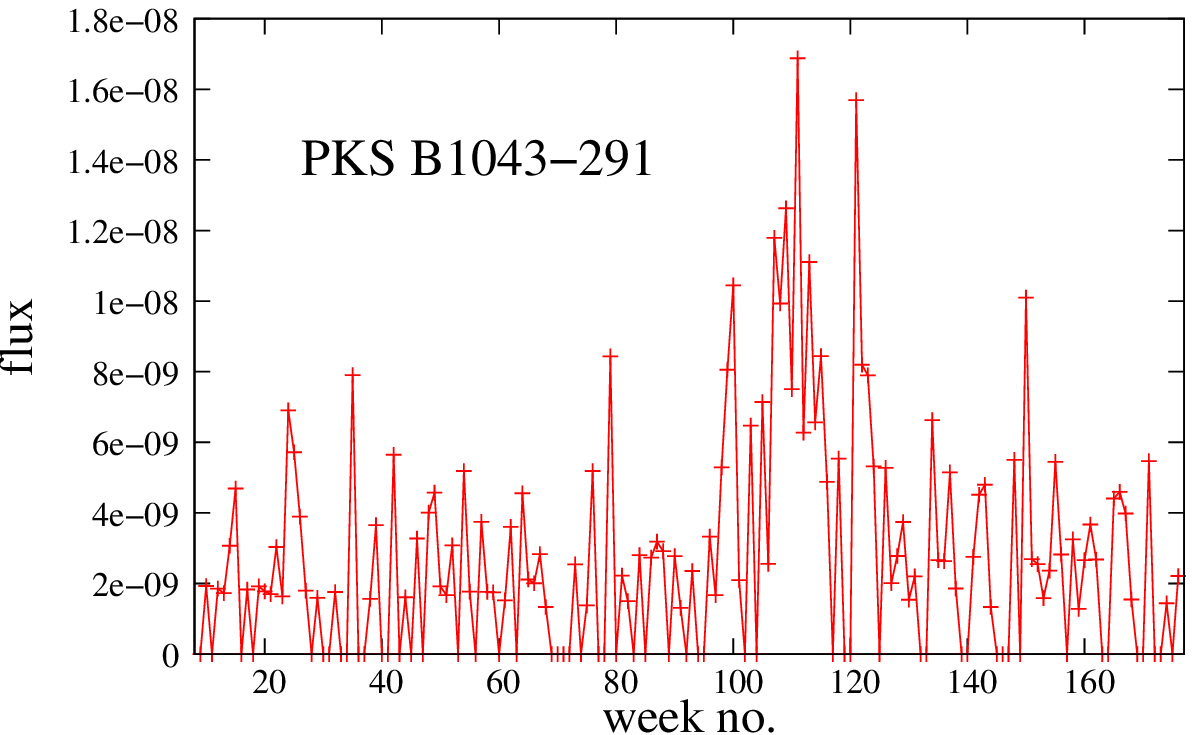}
\includegraphics[width=6.0cm]{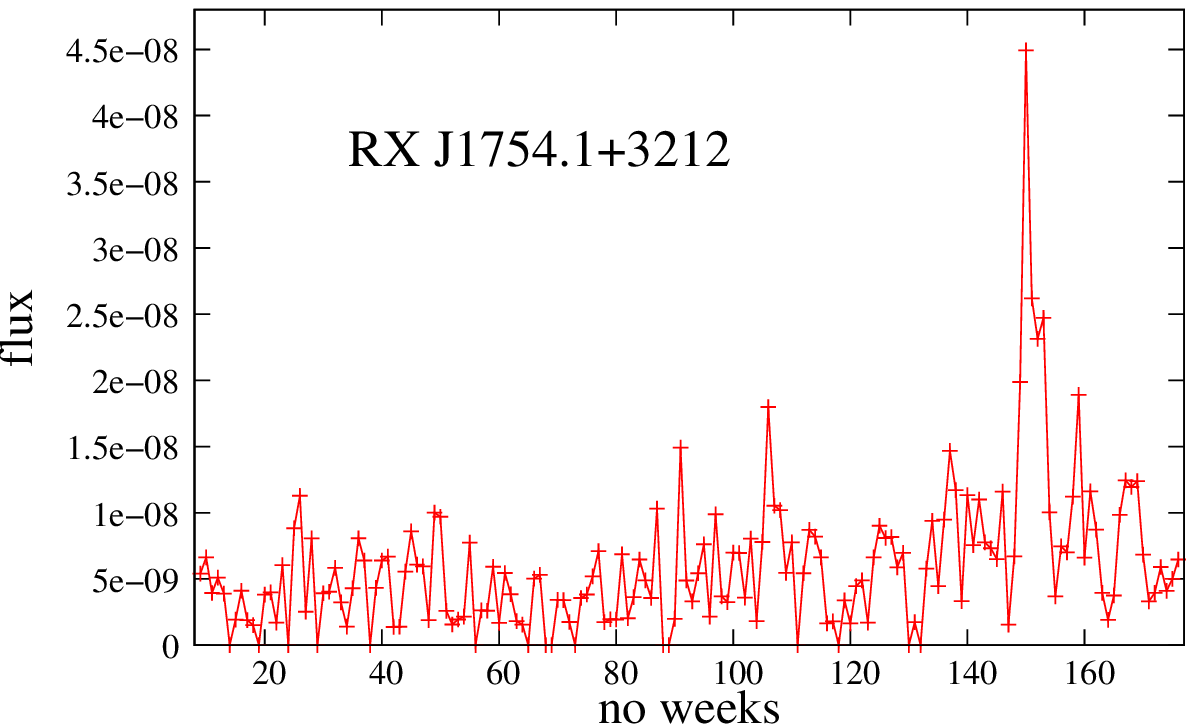}\\
\includegraphics[width=6.0cm]{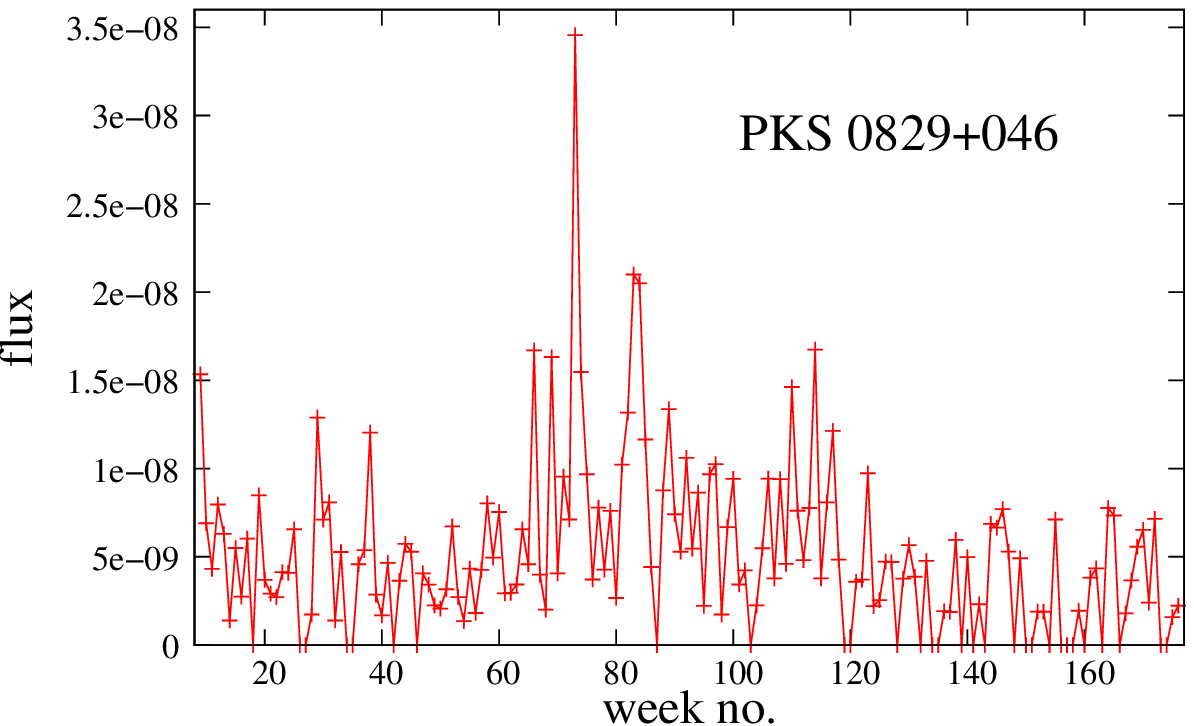}
\includegraphics[width=6.0cm]{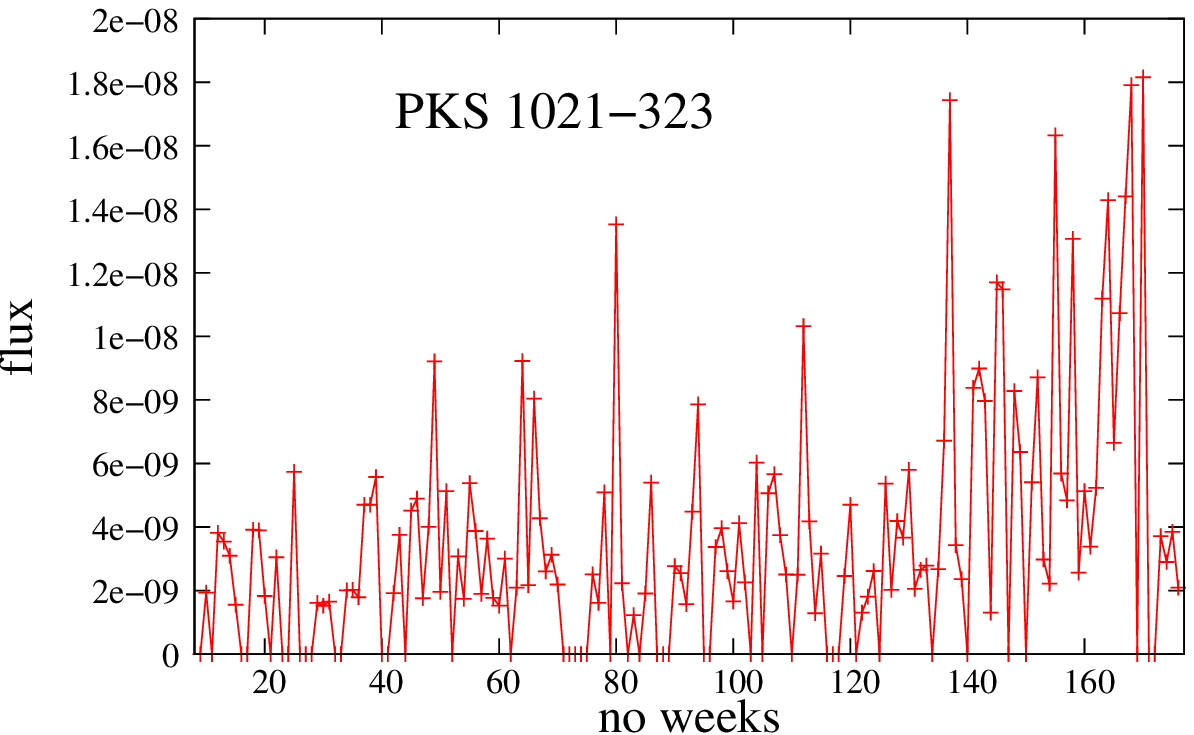}
\end{center}
\caption{The same as in Figure \ref{fig:flux1}. } \label{fig:flux2}
\end{figure}

\begin{figure}
\begin{center}
\includegraphics[width=6.0cm]{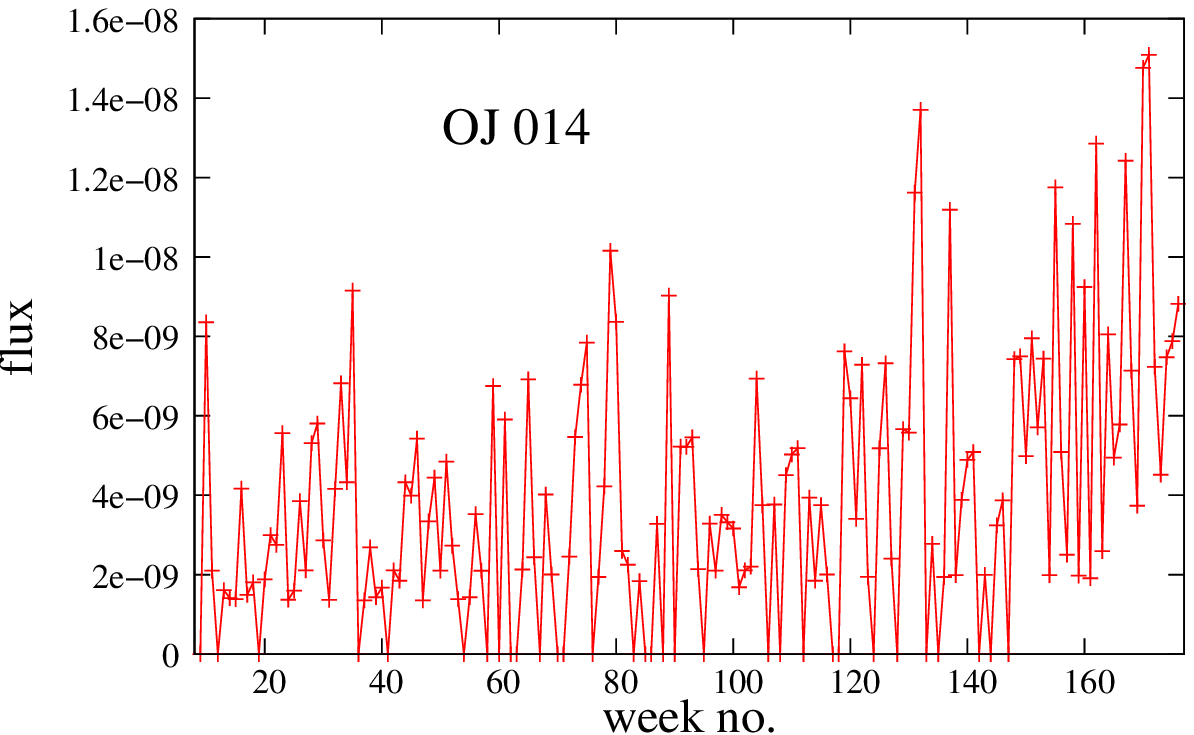}
\includegraphics[width=6.0cm]{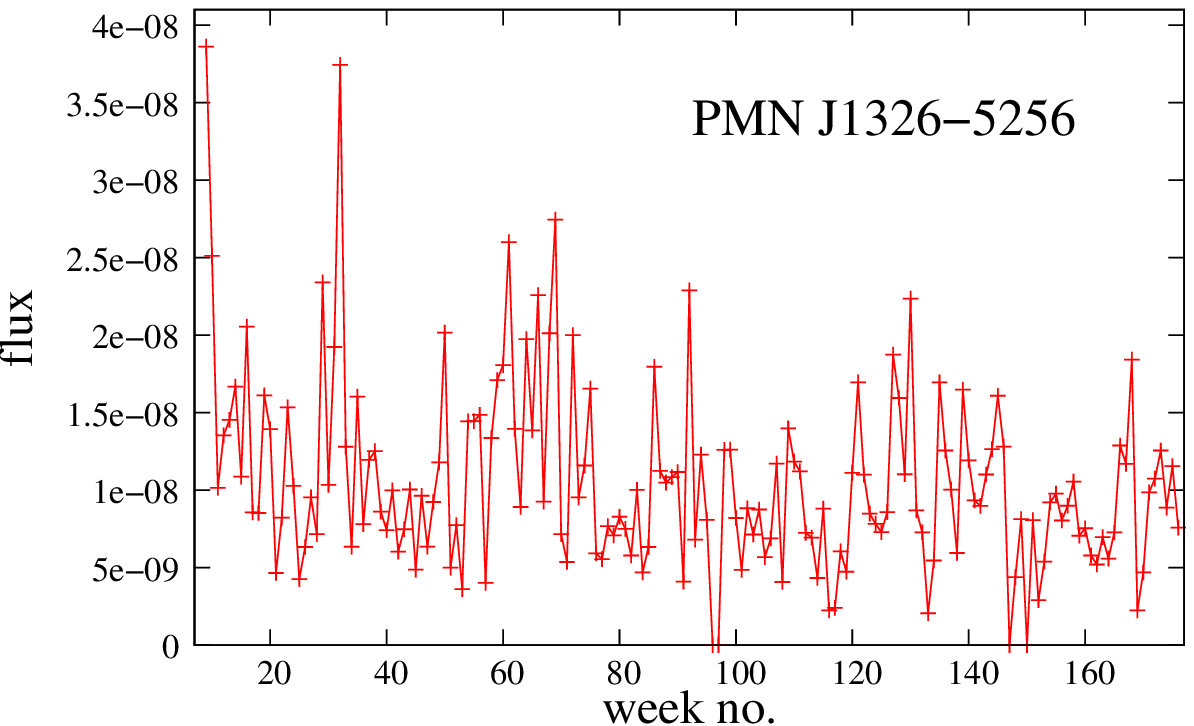}\\
\includegraphics[width=6.0cm]{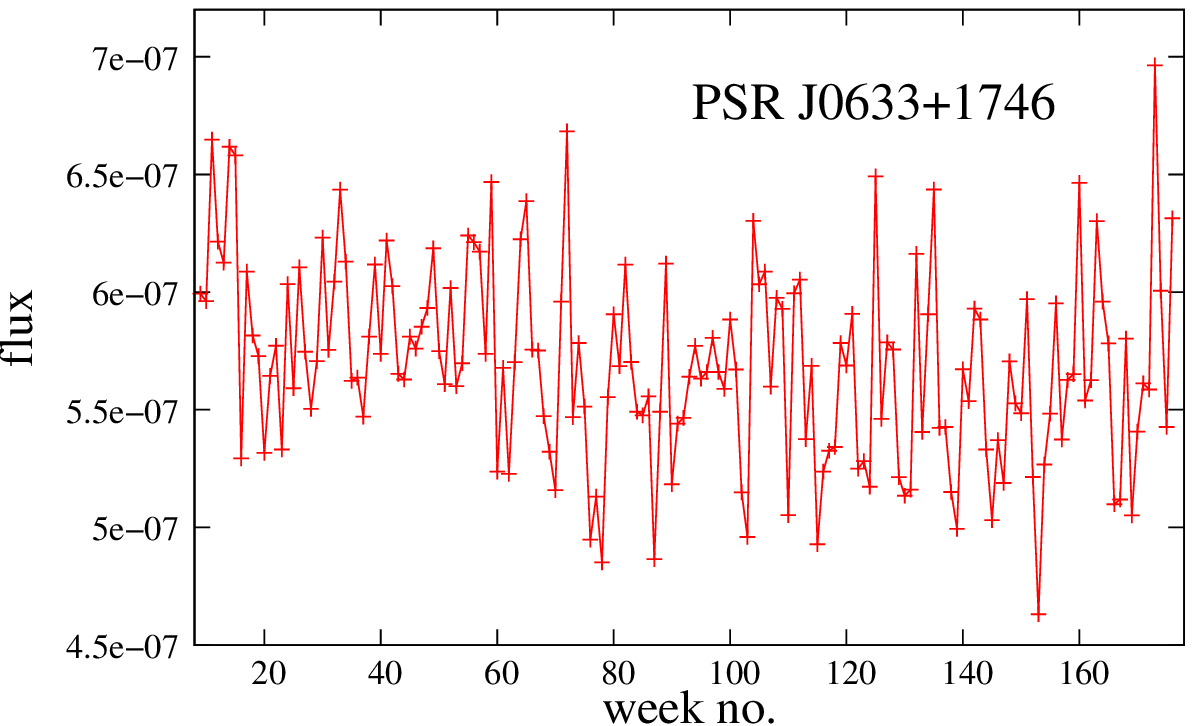}
\includegraphics[width=6.0cm]{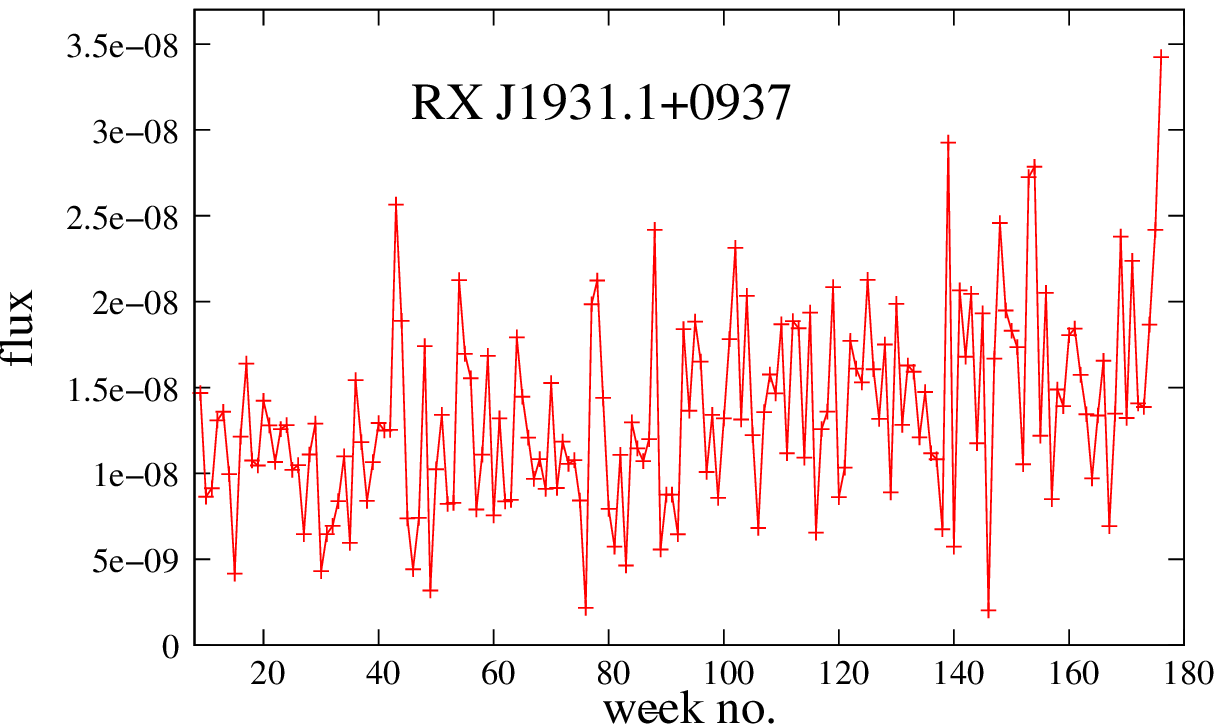}\\
\includegraphics[width=6.0cm]{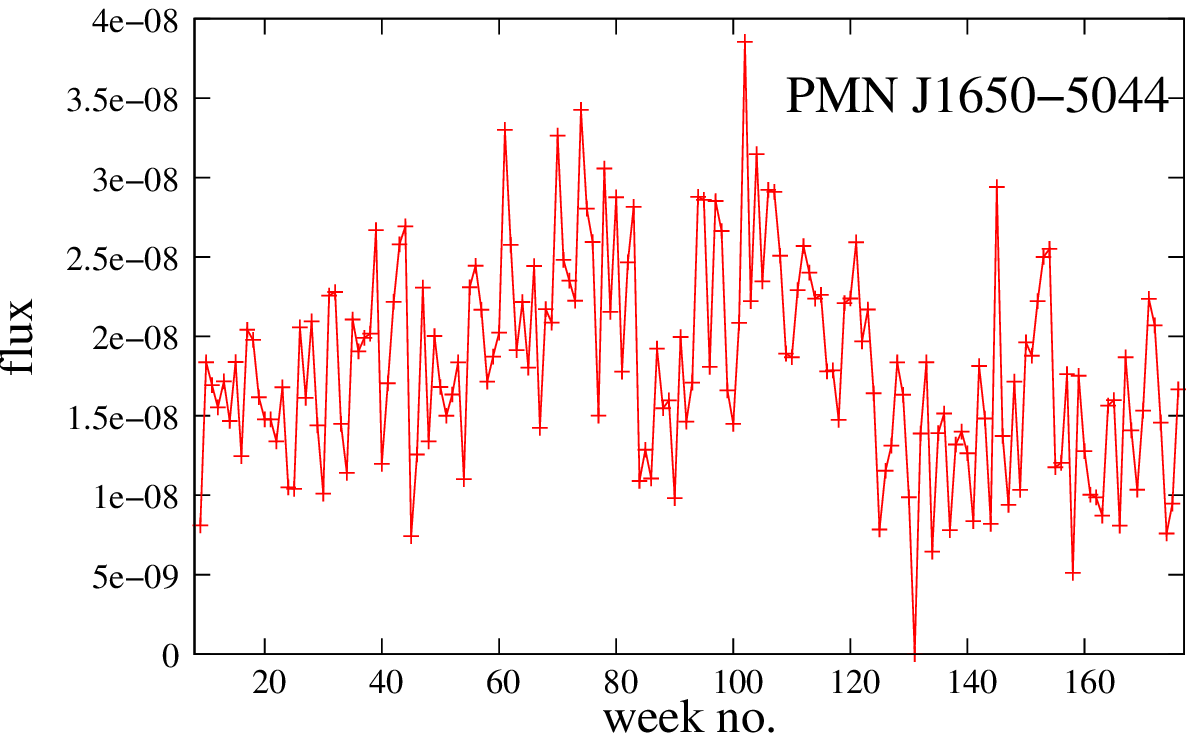}
\includegraphics[width=6.0cm]{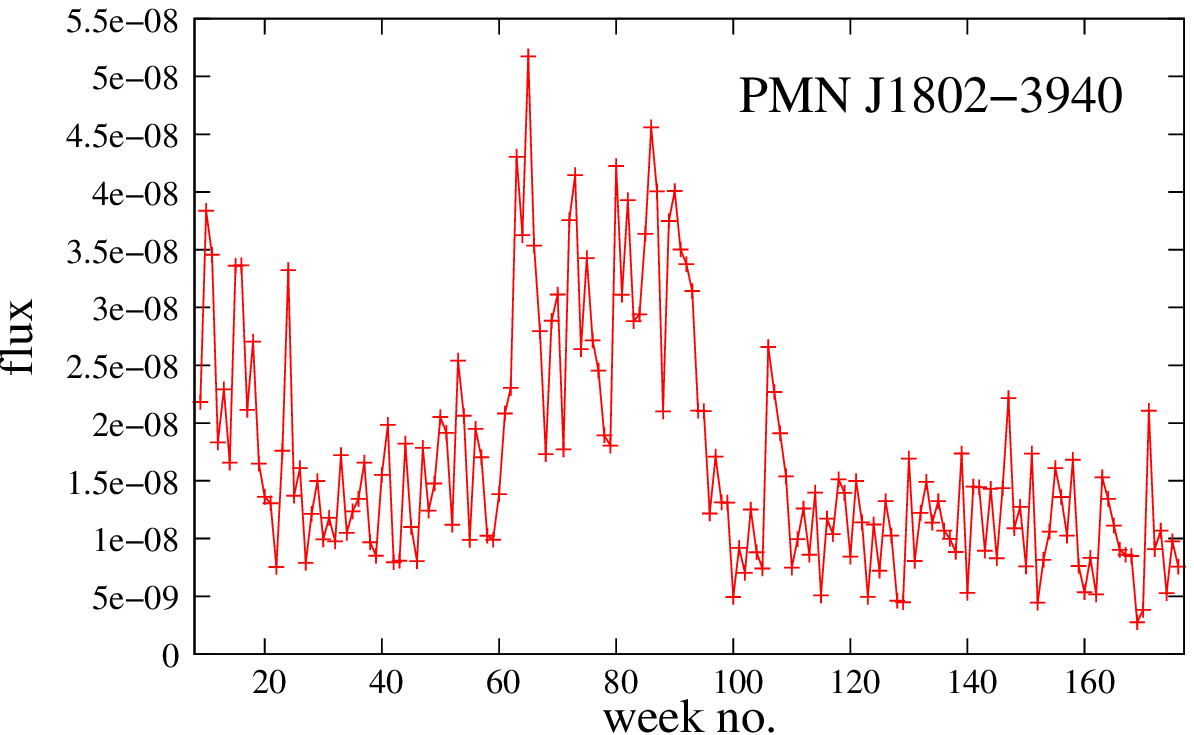}
\end{center}
\caption{The same as in Figure \ref{fig:flux1}.} \label{fig:flux3}
\end{figure}

\begin{figure}
\begin{center}
\includegraphics[width=6.0cm]{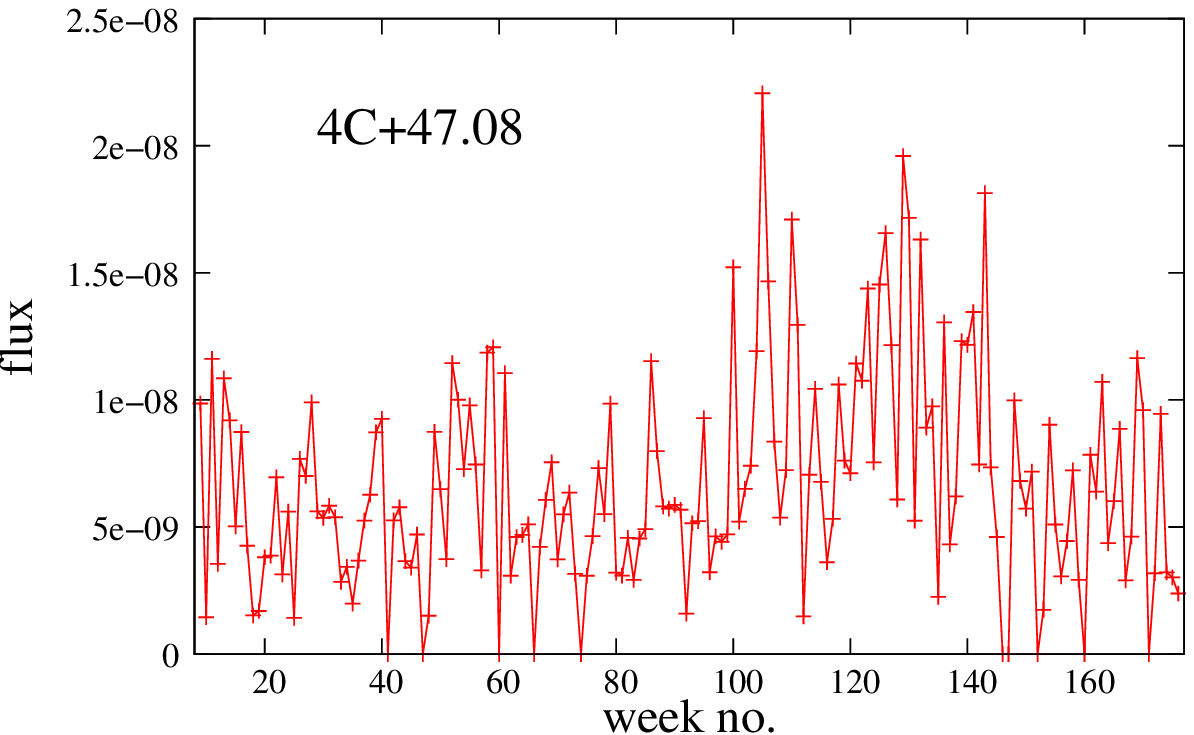}
\includegraphics[width=6.0cm]{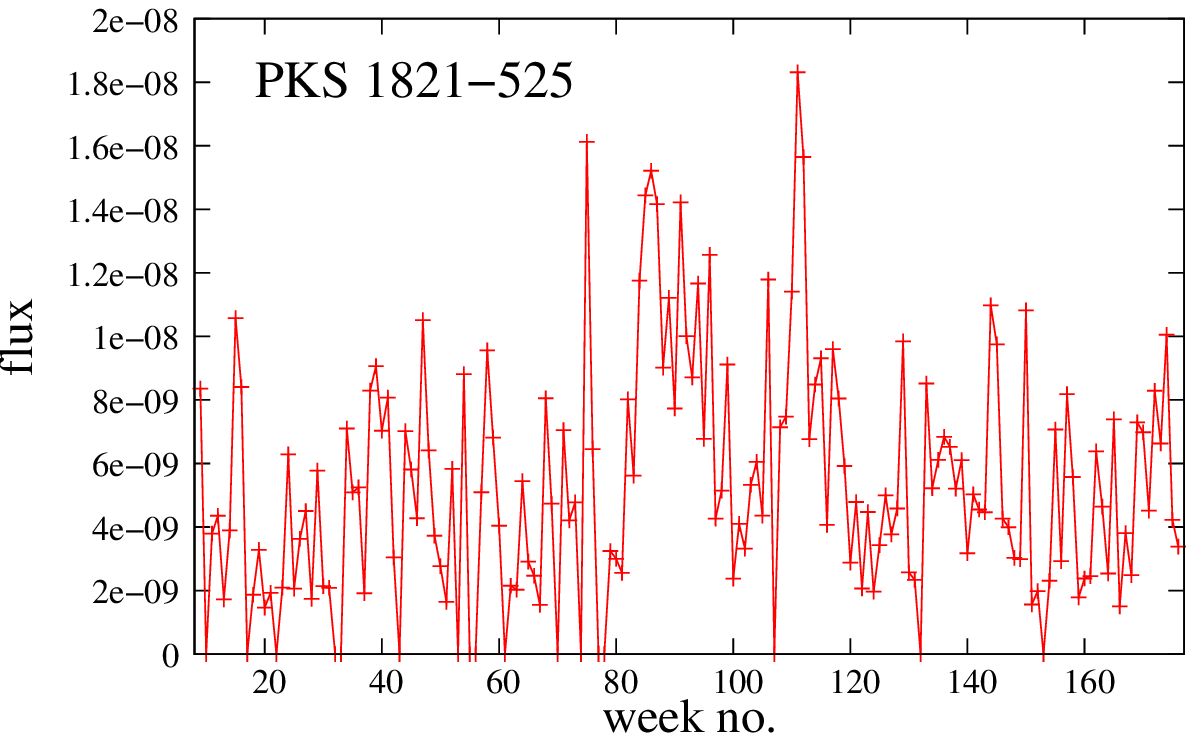}\\
\includegraphics[width=6.0cm]{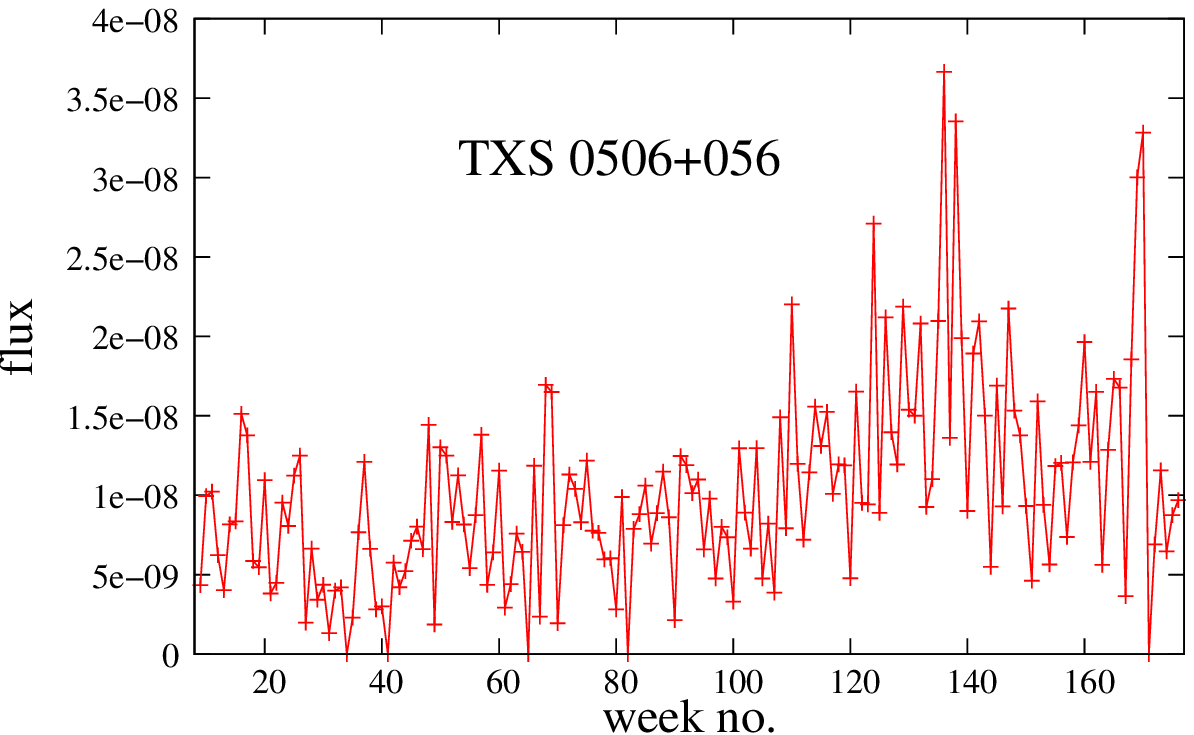}
\includegraphics[width=6.0cm]{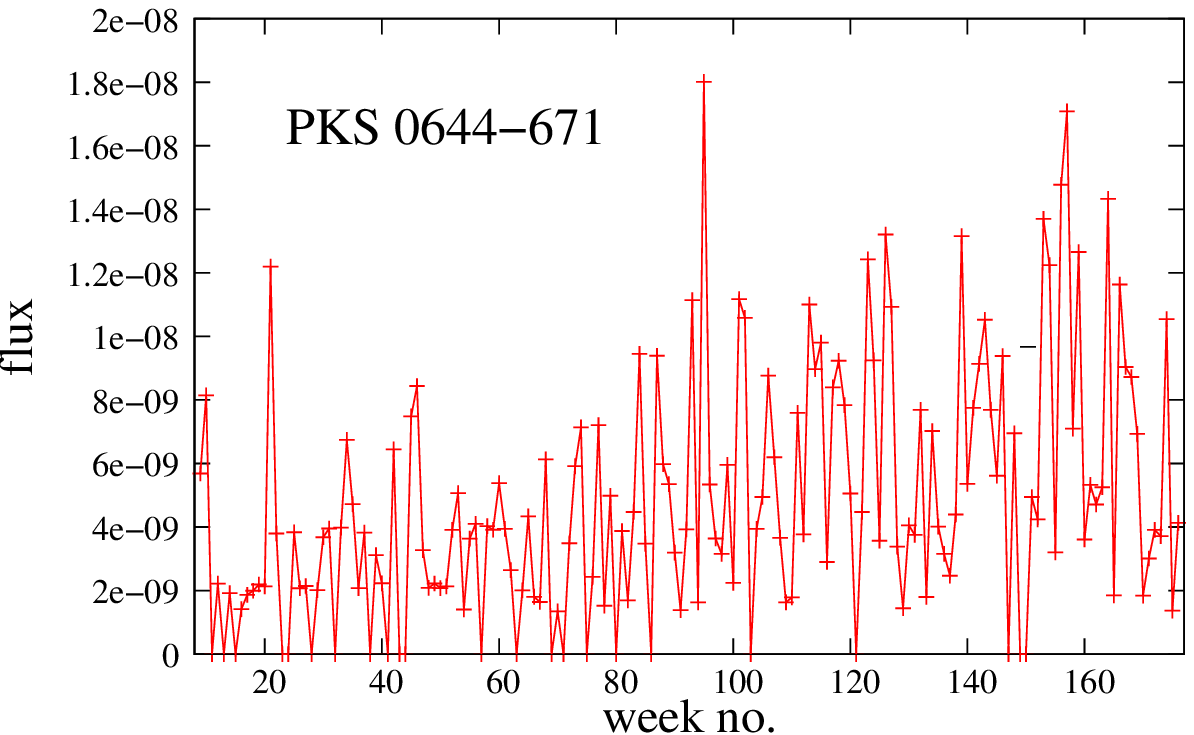}\\
\includegraphics[width=6.0cm]{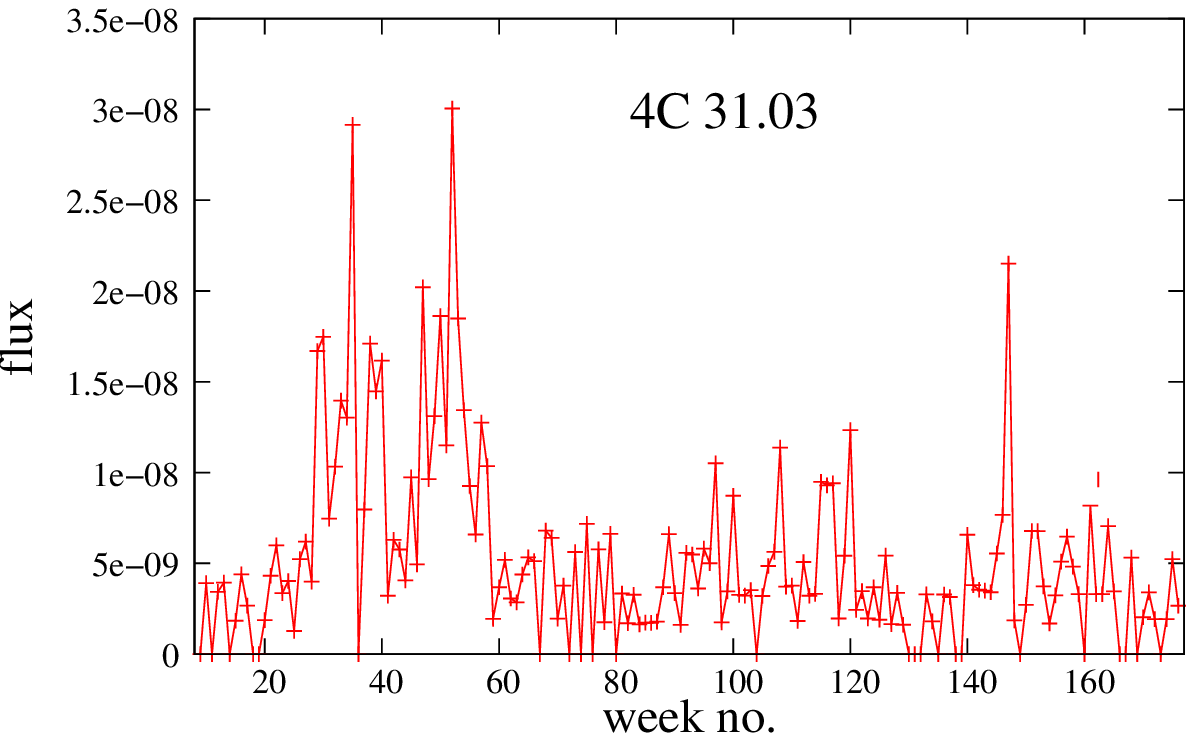}
\includegraphics[width=6.0cm]{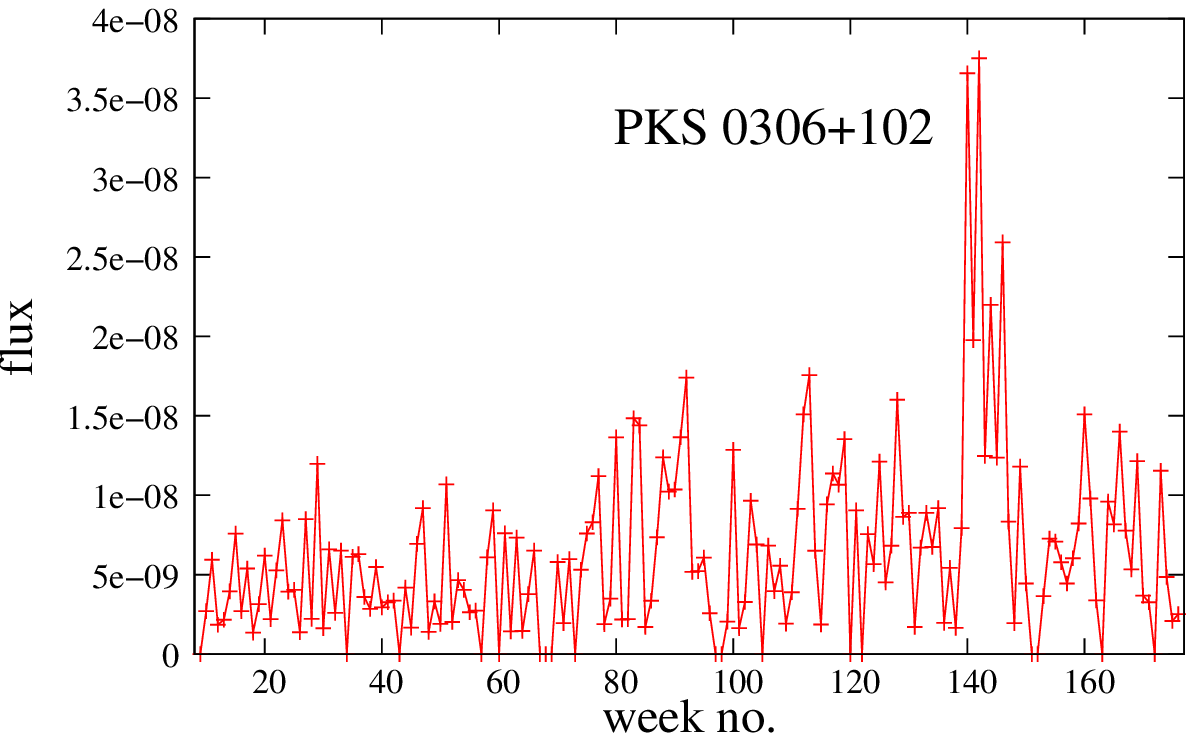}
\end{center}
\caption{The same as in Figure \ref{fig:flux1}.} \label{fig:flux4}
\end{figure}

\begin{figure}
\begin{center}
\includegraphics[width=6.0cm]{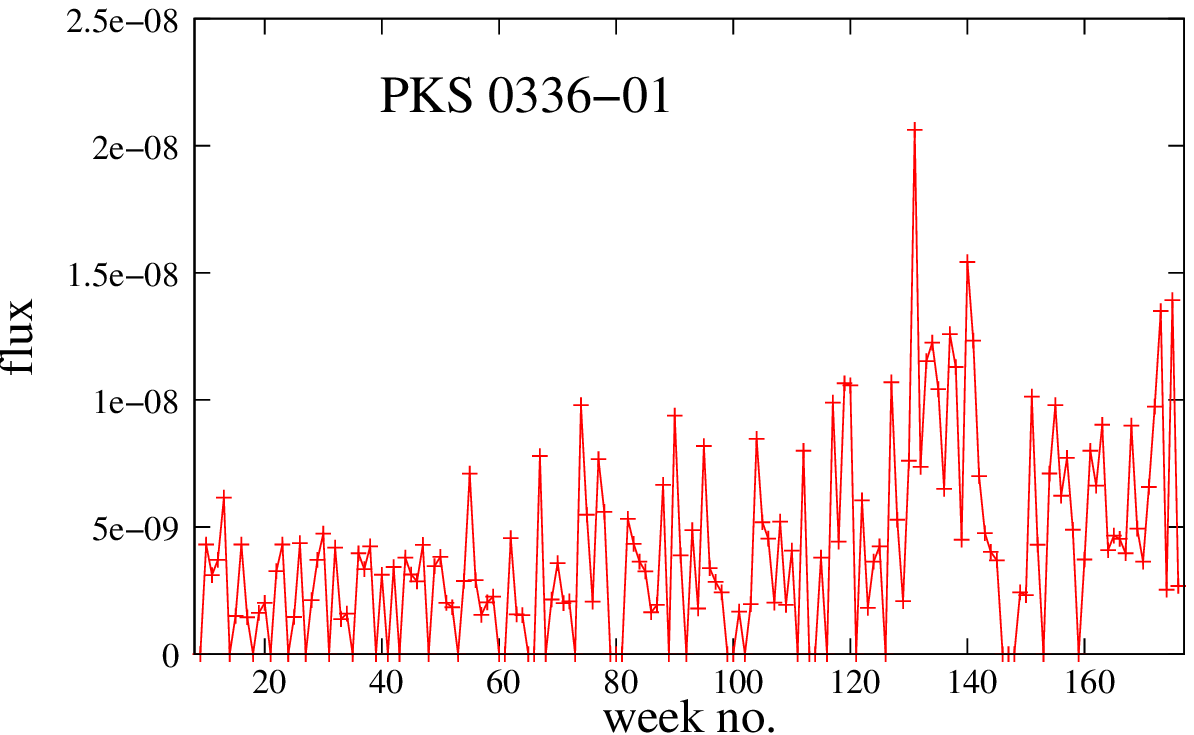}
\includegraphics[width=6.0cm]{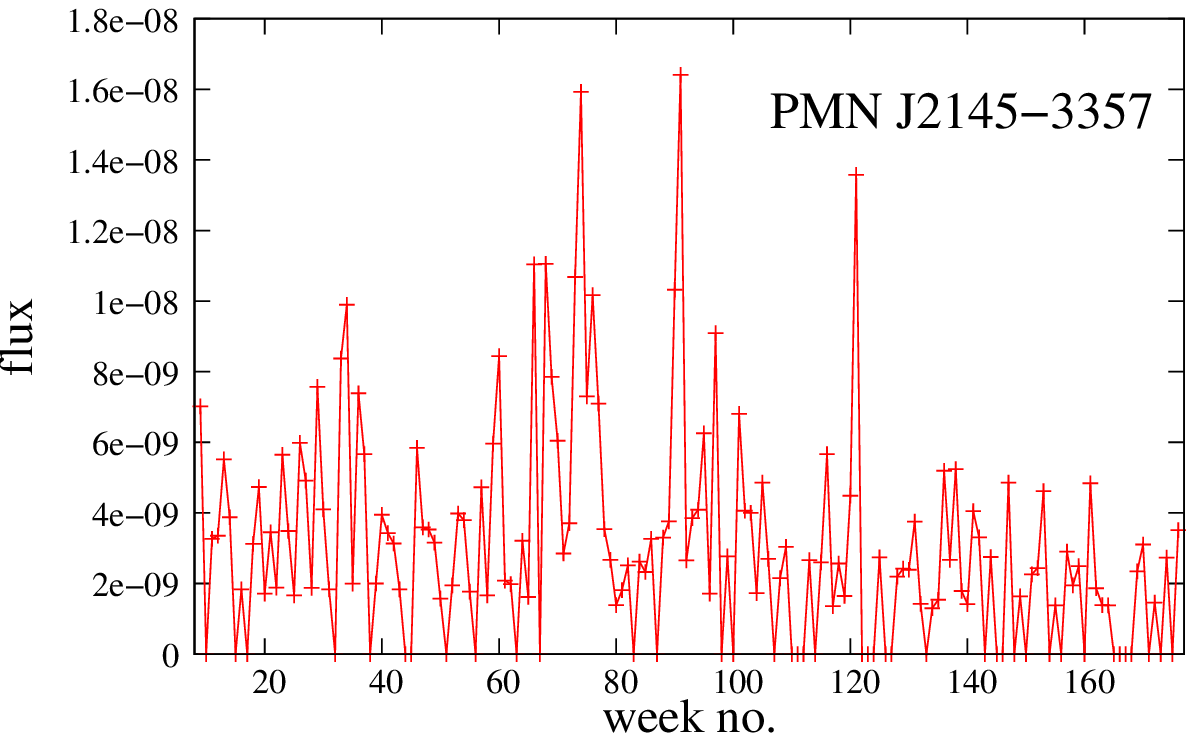}\\
\includegraphics[width=6.0cm]{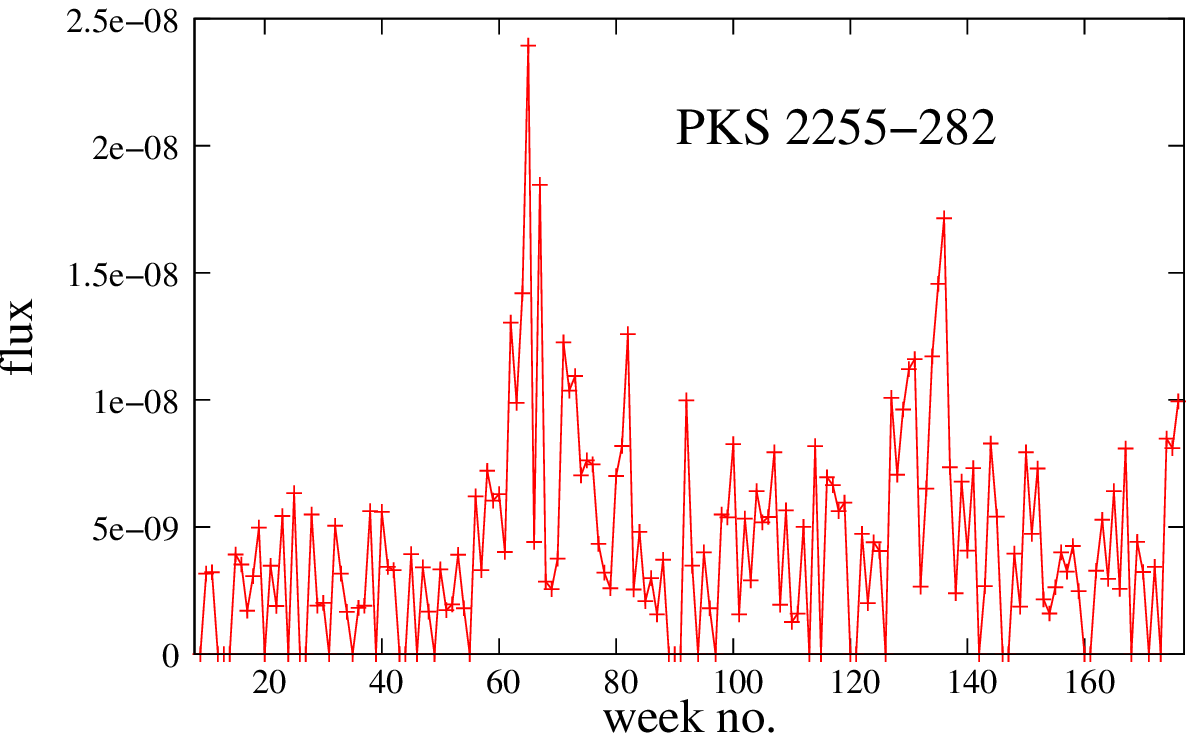}
\includegraphics[width=6.0cm]{empty.eps}
\end{center}
\caption{The same as in Figure \ref{fig:flux1}. } \label{fig:flux5}
\end{figure}

The total number of identified sources is 117; variability of 55 of
them was reported in \citep{Abdo2010a, Tanaka2011, Schinzel2011}, 35
additional detections were made in numerous Astronomers's telegrams
(ATels, see Appendix) and on the Fermi-LAT
blog \footnote{http://fermisky.blogspot.com/}. That leaves us with 27
sources  for which the variability has not been previously
  detected (see Table \ref{table:undetected} and Figures
\ref{fig:flux1}--\ref{fig:flux5}). We have explicitly checked that flux from these sources is not contaminated with the contribution from the Sun.

For convenience we assign a variability type to the source: a gradual
change in the photon flux is referred to as 'rate', if the whole
variability is dominated by one or several flares we call it 'flare'
and if these flares are observed by more or less prolonged time span
(typically, more than 20 weeks) we designate it as 'activity'. This
morphological distinction is not totally unambiguous: several bright
consequent flares could be defined as 'activity'. On the other hand,
gradual changes could take place on a time scales of several tens of
weeks, thus fitting the 'activity' type as well.

The BL Lacs and the flat-spectrum radio quasars (FSRQs) are
represented almost equally in the set of previously unidentified
sources: there are 11 BL Lacs, 14 FSRQs, and one AGN of uncertain type
(PKS 0644-671). Also there is one pulsar (Geminga) in the list (see
Table \ref{table:undetected} and Figures
\ref{fig:flux1}--\ref{fig:flux5}).  Both BL Lacs and FSRQs demonstrate
two types of variability: gradual change of photon flux (10 out of 25)
and flares or increased activity (15 out of 25). Variability of
several sources was observed before in other energy ranges: a flare in
the near infrared region was observed for B2 1732+38A (ATel \#3504
\citep{AT3504}, 1 July 2011), VHE flares of 1ES0806+524 were observed
by MAGIC (ATel \#3192 \citep{AT3192}, 24 February 2011), and EGRET
observed a very bright flare of PKS 2255-282 in December 1997
\citep{Macomb1999}.  Pulsed $\gamma$-rays from the Geminga pulsar were
observed with 1 year of Fermi-LAT data~\citep{Abdo2010b}, while the
source were considered non-variable. In this paper the long term flux
change of Geminga pulsar is detected with KS probability of $2.3\times
10^{-7}$. In present study Geminga is the only pulsar demonstrating
variability above our threshold. For comparison the KS probabilities
for bins containing Vela, PSR~J1709-4429, and Crab pulsars are 25\%,
78\% and 1.2\% respectively.  In Crab case it was shown that the
observed variability  is caused by processes in the Crab nebula rather
than the pulsar itself \citep{Buehler2011}. Fermi-LAT collaboration
also presented results of observations of three other gamma-ray
pulsars (J1836+5925, PSR J2021+4026, and PSR J0007+7303) where no flux
variability was observed \citep{Abdo2010c, SP2010,Abdo2012}. That
makes the case of Geminga even more intriguing.

We note that while 2FGL catalog contains 577 unidentified sources (out
of 1873), 153 of which have flux higher than $2\times 10^{-9}~
\mathrm{photons~cm^{-2}~s^{-1}}$, none of them show variability above
our threshold.

\begin{figure}
\begin{center}
\includegraphics[width=6.5 cm]{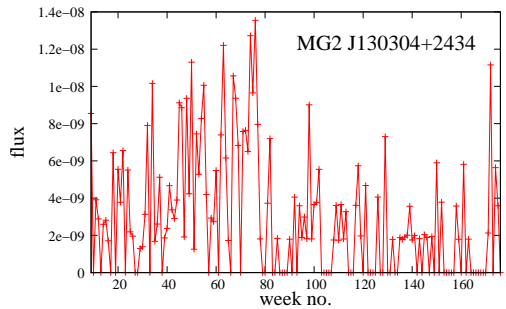}
\end{center}
\caption{Variability of MG2 J130304+2434. Change of rate could be easily seen.} \label{fig:flux_23}
\end{figure}

It is also worth noting that sources not included in the
Table~\ref{table:undetected} because of being reported either in ATels
or on the Fermi blog could have the variability patterns that differ
considerably from the reported one. As an example, the flare from the
source MG2 J130304+2434 that took place on 3 July 2009 (week no. 56)
was reported in ATel \#2110 \citep{AT2110}. On the other hand, Figure
\ref{fig:flux_23} shows that the flare occured during the high state
of the source, with its flux slowly increasing from the start of the
Fermi observations till approximately the 80-th week when it began to
decrease.

\section{Conclusions}
\label{sec:conclusions}
 A method for variable sources detection is proposed that uses the
KS statistical test. The method is implemented for a full sky blind
search for regions with variable flux at energies above 1 GeV using
Fermi-LAT 168 weeks data. The search leads to identification of 117
variable sources, the variability of 27 of which has not been reported
before. Among the sources with previously unidentified variability
there are 25 AGNs belonging to blazar class (11 BL Lacs and 14 FSRQs),
one AGN of uncertain type (PKS 0644-671), and one pulsar
PSR~J0633+1746~(Geminga).  The observed long term flux
  variability of Geminga pulsar has a statistical significance of
  5.1$\sigma$.

\section*{Appendix}
\label{appendix}

The following ATels are cited in the text: ATel \#1933 ~\citep{AT1933}, ATel \#2048 ~\citep{AT2048}, ATel \#2049 ~\citep{AT2049}, ATel \#2104 ~\citep{AT2104}, ATel \#2110 ~\citep{AT2110}, ATel \#2136 ~\citep{AT2136}, ATel \#2243 ~\citep{AT2243}, ATel \#2316 ~\citep{AT2316}, ATel \#2402 ~\citep{AT2402}, ATel \#2413 ~\citep{AT2413}, ATel \#2539 ~\citep{AT2539}, ATel \#2583 ~\citep{AT2583}, ATel \#2669 ~\citep{AT2669}, ATel \#2783 ~\citep{AT2783}, ATel \#2829 ~\citep{AT2829}, ATel \#2860 ~\citep{AT2860}, ATel \#2907 ~\citep{AT2907}, ATel \#2943 ~\citep{AT2943}, ATel \#3002 ~\citep{AT3002}, ATel \#3026 ~\citep{AT3026}, ATel \#3171 ~\citep{AT3171}, ATel \#3192 ~\citep{AT3192}, ATel \#3207 ~\citep{AT3207}, ATel \#3452 ~\citep{AT3452}, ATel \#3445 ~\citep{AT3445}, ATel \#3504 ~\citep{AT3504}, ATel \#3579 ~\citep{AT3579}, ATel \#3670 ~\citep{AT3670}, ATel \#3793 ~\citep{AT3793}.

\begin{acknowledgements}
 We are indebted to P.~Tinyakov for numerous helpful discussions at
 all stages of this work.  We thank M.~Gustafsson, B.~Stern,
   I.~Tkachev and S.~Troitsky for useful comments and suggestions.
 The work was supported in part by the RFBR grants 10-02-01406a,
 11-02-01528a, 12-02-91323-SIGa~(GR), by the grants of the President
 of the Russian Federation NS-5525.2010.2 (GR), MK-1632.2011.2 (GR),
 MK-1582.2010.2 (MP). The work of M.P is supported in part by the IISN
 project No.~4.4509.10 and the Belgian Science Policy (IAP VI-11). GR
 is grateful for the hospitality of ULB Service de Physique Theorique
 where this study was initiated. The analysis is based on data
   and software provided by the Fermi Science Support Center (FSSC).
 The numerical part of the work is performed at the cluster of the
 Theoretical Division of INR RAS. This research has made use of NASA's
 Astrophysics Data System and SIMBAD database, operated at CDS,
 Strasbourg, France.
\end{acknowledgements}

\bibliographystyle{aa}



\onecolumn
\LTcapwidth=\textwidth
\begin{center}
\begin{longtable}{|c|c|c|c|c|c|c|c|c|c|}
\hline no&Pixel no. & $l^{\circ}$ & $b^{\circ}$
&$N_{\mathrm{phot}}$ &$\Phi_{-8}  $& $P$ & source& 2FGL &ref
\endhead \hline
1&23 &  345.0  & 85.6 &  264& 0.29 &$2.8\times10^{-13}$   & MG2 J130304+2434 & J1303.1+2435 & AT2110\\
2&44 &  81.0  & 82.7 & 454& 0.48 &$6.5\times10^{-14}$   & OP 313 &J1310.6+3222 &P \\
3& 51& 207.0& 82.7& 467& 0.50& $1.4\times10^{-6}$& W Comae & J1221.4+2814 &P\\
4&54 &  261.0  & 82.7 &  1700& 1.86 &$4.3\times10^{-80}$   & 4C+21.35&J1224.9+2122&P2   \\
5& 66& 97.5& 81.2& 194& 0.20& $3.9\times10^{-6}$& 5C 12.291 & J1308.5+3547&\\
6&76 &  247.5  & 81.2 &  935& 1.02 &$2.4\times10^{-159}$   & 4C+21.35&J1224.9+2122 &P2  \\
7&77 &  262.5  & 81.2 &  475& 0.52 &$1.3\times10^{-27}$   & 4C+21.35&J1224.9+2122  &P2\\
8& 103& 250.7& 79.7& 216& 0.23& $1.4\times10^{-6}$& 4C +21.35& J1224.9+2122&P2 \\
9&129 &  196.9  & 78.3 &  783& 0.84 &$7.7\times10^{-40}$   & Ton 599 &J1159.5+2914 &P \\
10&335 &  162.7  & 70.9 &  468& 0.48 &$1.9\times10^{-16}$   & S4 1144+40 &J1146.9+4000 &B165  \\
&&&&&&&&&\\

11& 373& 61.1& 69.4& 188& 0.20&  $1.8\times10^{-9}$& J1424+3615& J1424+3615 &  \\
12& 378& 93.2& 69.4& 332& 0.33& $4.2\times10^{-10}$& B3 1343+451 & J1345.4+4453&AT3793 \\
13&381 &  112.5  & 69.4 &  433& 0.43 &$1.9\times10^{-11}$   & GB 1310+487 &J1312.8+4828 &AT2316 \\
14&438 &  111.0  & 67.9 &  343& 0.33 &$2.1\times10^{-22}$   & GB 1310+487  &J1312.8+4828 &AT2316\\
15& 532& 295.3& 66.4& 461& 0.54& $7.3\times10^{-9}$& MG1 J123931+0443& J1239.5+0443&AT3445 \\
16& 564& 108.5& 64.9& 201& 0.19& $1.4\times10^{-7}$& CLASS J1333+5057& J1333.5+5058&  \\
17&571 &  145.6  & 64.9 &  370& 0.36 &$7.5\times10^{-61}$   & OM 484&J1153.2+4935&B153 \\
18&598 &  288.5  & 64.9 &  855& 1.00 &$3.3\times10^{-56}$   & 3C 273 &J1229.1+0202 &P \\
19& 670& 292.5& 63.4& 301& 0.35& $2.1\times10^{-6}$& 3C 273& J1229.1+0202&P \\
20&998 &  304.8  & 57.4 &  1925& 2.26 &$5.1\times10^{-57}$   & 3C 279 &J1256.1-0547&P  \\
&&&&&&&&&\\

21&1014 &  9.8  & 55.9 &  967& 1.10 &$6.3\times10^{-123}$   & PKS 1502+106 &J1504.3+1029&P\\
22&1106 &  9.4  & 54.3 &  598& 0.68 &$2.0\times10^{-62}$   & PKS 1502+106 &J1504.3+1029&P\\
23&1107 &  13.1  & 54.3 &  1414& 1.60 &$4.1\times10^{-212}$   & PKS 1502+106&J1504.3+1029&P \\
24& 1159& 208.1& 54.3& 443& 0.48& $3.4\times10^{-11}$& MG2 J101241+2439& J1012.6+2440&P \\
25&1203 &  12.6  & 52.8 &  358& 0.41 &$4.1\times10^{-24}$   & PKS 1502+106&J1504.3+1029 &P\\
26&1448 &  148.3  & 49.7 &  744& 0.64 &$3.5\times10^{-60}$   & S4 1030+61 &J1033.9+6050 &P\\
27& 1483& 265.0& 49.7& 321& 0.38& $2.0\times10^{-10}$& PKS 1118-05 & J1121.5-0554&B71  \\
28& 1499& 318.3& 49.7& 310& 0.36& $8.1\times10^{-7}$& PMN J1332-1256 & J1332.5-1255&  \\
29&1585&    236.2&  48.1&   267&    0.31&   $3.0\times10^{-12}$   & PMN J1016+0512 &  J1016.0+0513&P\\
30& 1631& 23.3& 46.6& 374& 0.42& $7.6\times10^{-10}$& PKS 1551+130 & J1553.5+1255 &P\\

&&&&&&&&&\\

31& 1654& 94.7& 46.6& 465& 0.40& $1.4\times10^{-9}$& GB6 J1542+6129& J1542.9+6129&P \\
32&1700&    237.4&  46.6&   283&    0.33&   $3.3\times10^{-12}$   & PMN J1016+0512 &  J1016.0+0513&P\\
33&1798 &  175.5  & 45.0 &  729& 0.73 &$1.2\times10^{-44}$   & S4 0917+44&J0920.9+4441&P \\
34&1970 &  320.8  & 43.4 &  538& 0.62 &$8.3\times10^{-15}$   & PMN J1344-1723&J1344.2-1723&  \\
35&2005 &  60.5  & 41.8 &  1193& 1.23 &$2.7\times10^{-22}$   & 4C+38.41 &J1635.2+3810&P\\
36& 2006& 63.3& 41.8& 808& 0.82& $1.7\times10^{-6}$& NRAO 512 & J1640.7+3945&P3 \\
37&2237 &  351.6  & 40.2 &  2977& 3.47 &$1.5\times10^{-61}$   & PKS 1510-08 &J1512.8-0906&P\\
38&2299&    167.3&  38.7&   276&    0.26&   $8.1\times10^{-13}$  &SBS 0846+513 &  &AT3452\\
39&2338 &  277.0  & 38.7 &  527& 0.60 &$2.4\times10^{-33}$   & PKS 1124-186&J1126.6-1856 &AT3207\\
40& 2392& 67.5& 37.2& 254& 0.25& $1.1\times10^{-7}$& B3 1708+433 & J1709.7+4319&AT3026\\
&&&&&&&&&\\

41& 2475& 300.9& 37.2& 940& 1.06& $1.4\times10^{-11}$& PKS 1244-255& J1246.7-2546&P \\
42&2520 &  68.9  & 35.7 &  498& 0.50 &$1.9\times10^{-26}$   & B3 1708+433&J1709.7+4319 &AT3026\\
43& 2569& 206.7& 35.7& 551& 0.60& $8.9\times10^{-12}$&OJ 287 & J0854.8+2005&P \\
44& 2683& 165.9& 34.2& 258& 0.24& $4.7\times10^{-7}$& 1ES 0806+524& J0809.8+5218&VHE AT3192 \\
45&2749 &  351.6  & 34.2 &  628& 0.73 &$1.4\times10^{-77}$   & PMN J1532-1319 & &AT3579\\
46& 2810& 164.5& 32.8& 254& 0.24& $1.9\times10^{-7}$& 1ES 0806+524& J0809.8+5218&VHE AT3192 \\
47&2811 &  167.3  & 32.8 &  537& 0.51 &$9.6\times10^{-16}$   & 1ES 0806+524 &J0809.8+5218 &VHE AT3192\\
48& 2815& 178.6& 32.8& 735& 0.74&$1.2\times10^{-6}$& S4 0814+42&  J0818.2+4223&P \\
49& 2844& 260.2& 32.8& 290& 0.33& $1.3\times10^{-7}$& 1RXS J102658.5-174905&   J1026.7-1749&    \\
50&2877 &  353.0  & 32.8 &  431& 0.50 &$3.9\times10^{-15}$    & PMN J1532-1319 & &AT3579\\

&&&&&&&&&\\

51&2903 & 64.7  &31.4&  404&    0.41&   $2.1\times10^{-16}$ & B2 1732+38A&J1734.3+3858&IR AT3504\\
52&2916 &   101.2&  31.4&   466&    0.38&   $9.2\times10^{-21}$ & S4 1749+70&J1748.8+7006&AT3171\\
53& 3043& 99.8& 30.0& 830& 0.67& $4.0\times10^{-8}$&S4 1749+70 & J1748.8+7006&AT3171\\
54&3150 &  39.4  & 28.6 &  371& 0.40 &$1.9\times10^{-21}$   & PKS 1717+177 &J1719.3+1744 &P\\
55& 3187& 143.4& 28.6& 1845& 1.49& $4.0\times10^{-10}$& S5 0716+71 & J0721.9+7120&P \\
56&3194 &  163.1  & 28.6 &  756& 0.69 &$1.7\times10^{-50}$   & GB6 J0742+5444 &J0742.6+5442&AT3445\\
57& 3246& 309.4& 28.6& 451& 0.50& $2.7\times10^{-10}$& PKS 1313-333& J1315.9-3339&B122 \\
58& 3489& 272.8& 25.9& 247& 0.28& $1.3\times10^{-7}$& PKS B1043-291&J1045.5-2931&  \\
59&3540&    57.7&   24.6&   557&    0.58& $1.1\times10^{-16}$ & RX J1754.1+3212&J1754.3+3212&\\
60&3554 &  97.0  & 24.6 &  896& 0.73 &$1.1\times10^{-50}$   & S4 1849+67 &  J1849.4+6706&P \\

&&&&&&&&&\\

61& 3598& 220.8& 24.6& 439& 0.50& $1.4\times10^{-7}$& PKS 0829+046 &  J0831.9+0429&\\
62& 3963& 165.9& 20.7& 462& 0.44& $5.9\times10^{-9}$& GB6 J0654+5042 & J0654.5+5043&P \\
63&4000&    270.0&  20.7&   325&    0.36&   $1.7\times10^{-15}$ &  PKS 1021-323 & J1023.8-3248&\\
64&4021 &  321.1  & 20.7 &  813& 0.90 &$1.8\times10^{-44}$   & PKS 1454-354 & J1457.4-3540&P\\
65& 4092& 170.2& 19.5& 457& 0.45& $5.9\times10^{-10}$& B3 0650+453 &J0654.2+4514 &P \\
66& 4097& 184.2& 19.5& 613& 0.64& $9.7\times10^{-8}$& B2 0716+33 & J0719.3+3306 &P\\
67& 4110& 220.8& 19.5& 324& 0.37& $2.0\times10^{-8}$& OJ 014& J0811.4+0149 &\\
68& 4116& 237.7& 19.5& 412& 0.48& $5.7\times10^{-8}$& BZQ J0850-1213& J0850.2-1212&B102,114 \\
69& 4149& 330.5& 19.5& 463& 0.51& $2.6\times10^{-7}$& PKS 1454-354& J1457.4-3540&P \\
70&4402 &  322.0  & 17.0 &  1495& 1.63 &$7.8\times10^{-36}$   & PKS B1424-418 & J1428.0-4206&AT2104,2583 \\
&&&&&&&&&\\

71&4541&    351.6& 15.7&    1505&   1.69&   $1.5\times10^{-19}$ & PKS 1622-253 & J1625.7-2526&P\\
72& 4614& 198.3& 14.5& 338& 0.37& $1.2\times10^{-6}$&MG2 J071354+1934 & J0714.0+1933&P\\
73&4616&    203.9&  14.5&   799&    0.88&   $4.0\times10^{-19}$& 4C+14.23&J0725.3+1426 &AT2243\\
74& 4625& 229.2& 14.5& 402& 0.46& $1.1\times10^{-6}$&PKS 0805-07 & J0808.2-0750&AT2048,2136 \\
75&4742& 196.9& 13.2&   413&    0.44&   $5.7\times10^{-17}$& MG2 J071354+1934&J0714.0+1933&P\\
76& 4744& 202.5& 13.2& 353& 0.38& $1.9\times10^{-6}$&4C+14.23 &J0725.3+1426 &AT2243\\
77&4753& 227.8& 13.2&   707&    0.82&   $5.5\times10^{-16}$& PKS 0805-07&J0808.2-0750&AT2048,2136 \\
78&4881& 229.2& 12.0&   556&    0.64&   $3.4\times10^{-14}$& PKS 0805-07&J0808.2-0750&AT2048,2136 \\
79&4932&    11.2&   10.8&   1337&   1.54&   $2.7\times10^{-16}$& PKS 1730-13&J1733.1-1307&AT3002\\
80& 5119& 178.6& 9.6& 619& 0.64& $9.9\times10^{-8}$& B2 0619+33 &J0622.9+3326&AT2829  \\

&&&&&&&&&\\

81& 5165& 308.0& 9.6& 1011& 1.06& $2.6\times10^{-7}$& PMN J1326-5256&J1326.7-5254&  \\
82& 5637& 195.5& 4.8& 52455& 56.6& $2.3\times10^{-7}$& PSR J0633+1746& J0633.9+1746& \\
83&5248 &  180.0  & 8.4 &  1212& 1.26 &$1.1\times10^{-46}$   & B2 0619+33 & J0622.9+3326&AT2829 \\
84&5777&    227.8&  3.6&    2002&   2.3&    $3.2\times10^{-12}$   & PKS 0727-11& J0730.2-1141 &AT2860 \\
85& 6596& 12.7& -4.8& 1854& 2.1&  $4.6\times10^{-7}$& PKS 1830-211& J1833.6-2104 & AT2943\\
86& 6608& 46.4& -4.8& 1198& 1.3&  $3.6\times10^{-7}$& RX J1931.1+0937 & J1931.1+0938& \\
87& 6711& 336.1& -4.8& 1717& 1.8& $2.6\times10^{-9}$&PMN J1650-5044&  J1650.1-5044 &\\
88&6724&    11.2&   -6.0&   1945&   2.2&    $2.9\times10^{-15}$   & PKS 1830-211 &J1833.6-2104 &AT2943 \\
89&7101 &  351.6  & -8.4 &  1553& 1.71 &$5.4\times10^{-38}$   & PMN J1802-3940& J1802.6-3940&\\
90& 7155& 144.8& -9.6& 696& 0.67& $3.4\times10^{-7}$& 4C+47.08 &  J0303.5+4713& \\

&&&&&&&&&\\

91&7265 &  92.8  & -10.8 &  1664& 1.65 &$1.9\times10^{-46}$   & BL Lac &J2202.8+4216&P \\
92& 7413& 150.5& -12.0& 931& 0.93& $1.2\times10^{-8}$& NGC 1275&  J0319.8+4130&P\\
93& 7494& 16.9& -13.2& 868& 0.98& $1.2\times10^{-7}$& PKS B1908-201& J1911.1-2005&P  \\
94& 7542& 151.9& -13.2& 1202& 1.21& $6.7\times10^{-9}$&NGC 1275&  J0319.8+4130&P\\
95& 7569& 227.8& -13.2& 539& 0.61& $2.3\times10^{-8}$& PKS 0627-199 &J0629.3-2001&B174  \\
96&7662 &  130.8  & -14.5 &  868& 0.84 &$9.1\times10^{-23}$   & OC 457 & J0136.9+4751&P\\
97&7750 &  16.9  & -15.7 &  801& 0.91 &$7.5\times10^{-36}$   & TXS 1920-211 & J1923.5-2105&P \\
98& 7921& 139.2& -17.0& 2228& 2.2&  $3.6\times10^{-10}$& 3C 66A & J0222.6+4302&P \\
99& 7993& 341.7& -17.0& 512& 0.54&  $3.5\times10^{-6}$& PKS 1821-525&  J1825.1-5231& \\
100&8030&    84.4&   -18.2&  677&    0.70&$5.8\times10^{-17}$   & B2 2155+31& J2157.4+3129 &P\\

&&&&&&&&&\\

101&8197&    195.5&  -19.5&  896&    1.01&   $4.9\times10^{-20}$   & TXS 0506+056 & J0509.4+0542& \\
102&8525&    36.6&   -23.3&  431&    0.50&   $8.0\times10^{-16}$   & PKS 2023-07 & J2025.6-0736&P\\
103&8653 &  38.0  & -24.6 &  563& 0.65 &$1.3\times10^{-31}$   & PKS 2023-07 & J2025.6-0736&P\\
104&8675 &  99.8  & -24.6 &  658& 0.68 &$4.8\times10^{-23}$   & B2 2308+34&J2311.0+3425&AT2783\\
105&8867&    278.4&  -25.9&  508&    0.50&   $1.1\times10^{-13}$   & PKS 0644-671&J0644.2-6713&\\
106&9074&    140.6& -28.6&   428&    0.44&   $1.4\times10^{-17}$   & B2 0200+30&J0204.0+3045&B134\\
107&9077 &  149.1  & -28.6 &  821& 0.86 &$1.2\times10^{-31}$   & 4C+28.07&J0237.8+2846&AT3670\\
108&9094 &  196.9  & -28.6 &  466& 0.53 &$1.6\times10^{-22}$   & PKS 0440-00&J0442.7-0017&AT2049\\
109&9197&    128.0&  -30.0&  512&    0.53&   $1.1\times10^{-17}$   & 4C 31.03&J0112.8+3208&\\
110&9369 &  250.3  & -31.4 &  3416& 3.67 &$1.6\times10^{-90}$   & PKS 0537-441&J0538.8-4405&P \\
&&&&&&&&&\\

111& 9431& 66.1& -32.8& 395& 0.44& $1.1\times10^{-8}$& PKS 2144+092 &  J2147.3+093&P\\
112&9477&    195.5&  -32.8&  588& 0.67&  $1.9\times10^{-15}$   & PKS 0420-01&J0423.2-0120&AT2402 \\
113&9498 &  254.5  & -32.8 &  946& 1.01 &$7.7\times10^{-101}$   & PMN J0531-4827 &J0532.0-4826 &AT2907 \\
114& 9615& 222.2& -34.2& 259& 0.30& $3.6\times10^{-8}$& PKS 0454-234& J0457.0-2325 &P\\
115&9616&    225.0&  -34.2&  341&    0.38&   $7.5\times10^{-14}$   &PKS 0454-234&J0457.0-2325 &P  \\
116&9743 &  223.6  & -35.7 &  1282& 1.43 &$5.8\times10^{-118}$   & PKS 0454-234&J0457.0-2325 &P\\
117&9776&    316.4&  -35.7&  483&    0.50&   $2.4\times10^{-17}$   & PKS 2142-75 &J2147.4-7534&AT2539\\
118&9822 &  84.4  & -37.2 &  782& 0.85 &$5.2\times10^{-23}$   &  3C 454.3 & J2253.9+1609&P\\
119&9823 &  87.2  & -37.2 &  1832& 1.98 &$1.5\times10^{-72}$   &  3C 454.3 &J2253.9+1609&P\\
120&9904&    315.0&  -37.2&  530&    0.55&   $1.8\times10^{-15}$   &   PKS 2142-75&J2147.4-7534&AT2539\\

&&&&&&&&&\\

121&9947&    77.3&   -38.7&  500&    0.55&   $4.2\times10^{-13}$   &  CTA 102 &J2232.4+1143&P\\
122& 9951& 88.6& -38.7& 417& 0.45& $1.8\times10^{-6}$& 3C 454.3& J2253.9+1609&P \\
123&9950 &  85.8  & -38.7 &  10048& 11.0 &$7.4\times10^{-539}$   &  3C 454.3&J2253.9+1609&P\\
124&9975 &  156.1  & -38.7 &  1097& 1.19 &$7.5\times10^{-247}$   &  AO 0235+164&J0238.7+1637&P\\
125& 10094& 129.4& -40.2& 838& 0.89& $4.4\times10^{-8}$& S2 0109+22 & J0112.1+2245&P \\
126&10104 &  157.5  & -40.2 &  522& 0.57 &$4.0\times10^{-37}$   &  AO 0235+164&J0238.7+1637&P\\
127&10108&   168.8&  -40.2&  564&    0.62&   $1.8\times10^{-12}$   &  PKS 0306+102&J0309.1+1027&\\
128& 10173& 351.6& -40.2& 648& 0.70& $8.4\times10^{-11}$& PKS 2052-47 &J2056.2-4715&P  \\
129&10368&   187.3&  -43.4&  339&    0.39&   $4.1\times10^{-15}$   &  PKS 0336-01&J0339.4-0144&\\
130&10386 &  239.5  & -43.4 &  1411& 1.54 &$1.8\times10^{-50}$   &  PKS 0426-380&J0428.6-3756&P\\

&&&&&&&&&\\

131& 10387& 242.4& -43.4& 440& 0.48& $8.9\times10^{-7}$& PKS 0426-380 &  J0428.6-3756&P\\
132&10508&   241.5&  -45.0&  683&    0.75&$1.3\times10^{-6}$   &  PKS 0426-380&J0428.6-3756&P\\
133& 10692& 91.6& -48.1& 289& 0.32& $4.8\times10^{-9}$& PKS 2325+093&J2327.5+0940&P \\
134&10711 &  152.7  & -48.1 &  569& 0.63 &$6.2\times10^{-22}$   &  MG1 J021114+1051&J0211.2+1050&P\\
135&10775&   358.4&  -48.1&  582&    0.64&   $1.1\times10^{-12}$   &  MH 2136-428&J2139.3-4236&P\\
136& 10779& 11.7& -49.7& 263& 0.29& $1.2\times10^{-7}$& PMN J2145-3357&  J2144.8-3356&\\
137& 10840& 215.0& -49.7& 186& 0.21& $1.1\times10^{-7}$& PKS 0347-211 &  J0350.0-2104&P\\
138& 10847& 238.3& -49.7& 337& 0.37& $3.5\times10^{-6}$& PKS 0402-362 &  J0403.9-3604&AT2413\\
139& 10889& 19.0& -51.3& 1101& 1.24& $6.0\times10^{-8}$& PKS 2155-304 & J2158.8-3013&P \\
140& 10965& 282.1& -51.3& 294& 0.29& $2.9\times10^{-7}$& PKS 0235-618 &  J0237.1-6136&AT2669\\

&&&&&&&&&\\

141& 10992& 16.2& -52.8& 1217& 1.36& $2.2\times10^{-7}$&PKS 2155-304 & J2158.8-3013 &P\\
142&11131&   163.1&  -54.3&  454&    0.52&   $7.3\times10^{-17}$   & PKS 0215+015&J0217.9+0143&P\\
143& 11495& 213.7& -60.4& 572& 0.64& $5.2\times10^{-11}$&PKS 0301-243 &  J0303.4-2407&P\\
144& 11506& 263.2& -60.4& 470& 0.51& $9.1\times10^{-8}$&PKS 0244-470 & J0245.9-4652 &P\\
145& 11572& 210.8& -61.9& 563& 0.63& $1.9\times10^{-9}$&PKS 0250-225& J0252.7-2218 &AT1933\\
146&11586&   277.1&  -61.9&  625&    0.67&   $3.5\times10^{-13}$   & PKS 0208-512& J0210.7-5102&P\\
147& 11597& 329.2& -61.9& 267& 0.28& $3.2\times10^{-11}$&PKS 2326-502 & J2329.2-4956&AT2783\\
148&11598 &  333.9  & -61.9 &  508& 0.55 &$1.5\times10^{-27}$   &  PKS 2326-502&J2329.2-4956&AT2783\\
149&11670 &  332.5  & -63.5 &  805& 0.87 &$9.0\times10^{-62}$   &  PKS 2326-502&J2329.2-4956&AT2783\\
150& 11680& 23.8& -64.9& 387& 0.43& $2.1\times10^{-7}$& PKS 2255-282 & J2258.0-2759 &\\
&&&&&&&&&\\

151&11933 &  65.8  & -70.9 &  953& 1.09 &$1.4\times10^{-90}$   &  PMN J2345-1555&J2345.0-1553&P \\

\hline
\caption{
List of pixels demonstrating  variability exceeding
threshold value ($P<4\times10^{-6}$). $l, b$ are the galactic
coordinates of the center of the pixel, $N_{\mathrm{phot}}$ is the total
number of photons observed in the pixel, $\Phi_{-8}$ is the
average flux from the pixel
$\Phi_{-8}\equiv~\Phi/{10^{-8}~\mathrm{photons~ cm^{-2}~
s^{-1}}}$: the total number of photons divided by the total
exposure, $P$ is the KS probability,  the designations    of  the
identified source in the literature and in the 2FGL catalog  are in the
8th and 9th columns. Previous references to the variability of the
source are presented in the last column: P stands for paper
\citep{Abdo2010a}, P2 for \citep{Tanaka2011}, P3 for \citep{Schinzel2011}, ATNNNN for ATel
\#NNNN, and BNNN indicates that the outburst from the source was
mentioned on the Fermi blog in the NNNth weekly report. ATNNNN with prefix VHE or IR indicate that flare was observed (and reported in corresponding ATel) in some other energy range: very high energy (larger than 100 GeV) or in the infrared. All the references for the ATels are listed in the Appendix.}

\label{table:high}
\end{longtable}
\end{center}

\begin{table}
\begin{center}
\begin{tabular}{|c|c|c|c|c|c|c|c|c|c|c|}
\hline
no&Pixel no. & $l^{\circ}$ & $b^{\circ}$ & $N_{\mathrm{phot}}$ &$\Phi_{-8}  $& $P$ & source & source type&variability& weeks  \\
\hline

1& 66& 97.5& 81.2& 194& 0.20& $3.9\times10^{-6}$& 5C 12.291 & Q&A &30--60\\
2& 373& 61.1& 69.4& 188& 0.20&  $1.8\times10^{-9}$& J1424+3615 &B &A &140--160 \\
3& 564& 108.5& 64.9& 201& 0.19& $1.4\times10^{-7}$& CLASS J1333+5057& Q&R &- \\
4& 1499& 318.3& 49.7& 310& 0.36& $8.1\times10^{-7}$& PMN J1332-1256 &Q &R &- \\
5&1970 &  320.8  & 43.4 &  538& 0.62 &$8.3\times10^{-15}$   & PMN J1344-1723& Q & R&-\\

6&2811 &  167.3  & 32.8 &  537& 0.51 &$9.6\times10^{-16}$   & 1ES 0806+524 &B& R&- \\
7& 2844& 260.2& 32.8& 290& 0.33& $1.3\times10^{-7}$& 1RXS J102658.5-174905&B   &R,F &176   \\

8&2903 & 64.7   &31.4&  404&    0.41&   $2.1\times10^{-16}$ & B2 1732+38A&B&F&40--50\\

9& 3489& 272.8& 25.9& 247& 0.28& $1.3\times10^{-7}$& PKS B1043-291&Q&A &100--120  \\
10& 3540& 57.7& 24.6& 557& 0.58& $1.1\times10^{-16}$& RX J1754.1+3212&B&F &149--153  \\
&&&&&&&&&&\\

11& 3598& 220.8& 24.6& 439& 0.50& $1.4\times10^{-7}$& PKS 0829+046 & B &F&73\\
12& 4000& 270.0& 20.7& 325& 0.36& $1.7\times10^{-15}$& PKS 1021-323 &Q&R &-\\
13& 4110& 220.8& 19.5& 324& 0.37& $2.0\times10^{-8}$& OJ 014& B&R &-\\
14& 5165& 308.0& 9.6& 1011& 1.06& $2.6\times10^{-7}$& PMN J1326-5256&B&F&9,32  \\
15& 5637& 195.5& 4.8& 52455& 56.6& $2.3\times10^{-7}$& PSR J0633+1746& PSR&R&- \\
16& 6608& 46.4& -4.8& 1198& 1.3&  $3.6\times10^{-7}$& RX J1931.1+0937 &B&R &- \\
17& 6711& 336.1& -4.8& 1717& 1.8& $2.6\times10^{-9}$&PMN J1650-5044&  Q&R &-\\
18&7101 &  351.6  & -8.4 &  1553& 1.71 &$5.4\times10^{-38}$   & PMN J1802-3940& Q&A& 60-100\\

19& 7155& 144.8& -9.6& 696& 0.67& $3.4\times10^{-7}$& 4C+47.08 & B &A&100-140 \\
20& 7993& 341.7& -17.0& 512& 0.54&  $3.5\times10^{-6}$& PKS 1821-525&  Q&R&- \\
&&&&&&&&&&\\

21&8197& 195.5&  -19.5&  896&    1.01&   $4.9\times10^{-20}$   & TXS 0506+056 & B& A&137-140, 169-172\\
22&8867& 278.4&  -25.9&  508&    0.50&   $1.1\times10^{-13}$   & PKS
0644-671& AGN& R& -\\
23&9197&    128.0&  -30.0&  512&    0.53&   $1.1\times10^{-17}$   & 4C 31.03&Q&A&30-50,148\\
24&10108&   168.8&  -40.2&  564&    0.62&   $1.8\times10^{-12}$   &  PKS 0306+102&Q&F&141-147\\
25&10368&   187.3&  -43.4&  339&    0.39&   $4.1\times10^{-15}$   &  PKS 0336-01&Q&F&131\\
26& 10779& 11.7& -49.7& 263& 0.29& $1.2\times10^{-7}$& PMN J2145-3357& Q &A& 74,91,121\\
27& 11680& 23.8& -64.9& 387& 0.43& $2.1\times10^{-7}$& PKS 2255-282 &Q & A&60--80, 136\\

 \hline
\end{tabular}
\end{center}
\caption{List of sources with previously unreported variability.
  Additional columns describe source type: BL Lac (B), FSRQ (Q),
  AGN of uncertain type (AGN), or pulsar (PSR), variability type: gradual
  increase or decrease in photon flux rate (R), flares (F) or longer
  period of increased activity (A); in case of flares or activity the
  temporal localization of events is given in the last
  column. }\label{table:undetected}
\end{table}

\newpage

\begin{thebibliography}{51}
\expandafter\ifx\csname
natexlab\endcsname\relax\def\natexlab#1{#1}\fi

\bibitem[{{Abdo} {et~al.}(2010{\natexlab{a}}){Abdo}, {Ackermann}, {Ajello},
  {Antolini}, {Baldini}, {Ballet}, {Barbiellini}, {Bastieri}, {Bechtol}, \&
  {Bellazzini}}]{Abdo2010a}
{Abdo}, A.~A., {Ackermann}, M., {Ajello}, M., {et~al.}
2010{\natexlab{a}},
  \apj, 722, 520

\bibitem[{{Abdo} {et~al.}(2010{\natexlab{b}}){Abdo}, {Ackermann}, {Ajello},
  {Atwood}, {Baldini}, {Ballet}, {Barbiellini}, {Baring}, {Bastieri}, \&
  {Bechtol}}]{Abdo2010c}
{Abdo}, A.~A., {Ackermann}, M., {Ajello}, M., {et~al.}
2010{\natexlab{b}},
  \apj, 712, 1209

\bibitem[{{Abdo} {et~al.}(2010{\natexlab{c}}){Abdo}, {Ackermann}, {Ajello},
  {Baldini}, {Ballet}, {Barbiellini}, {Bastieri}, {Baughman}, {Bechtol}, \&
  {Bellazzini}}]{Abdo2010b}
{Abdo}, A.~A., {Ackermann}, M., {Ajello}, M., {et~al.}
2010{\natexlab{c}},
  \apj, 720, 272

\bibitem[{{Abdo} {et~al.}(2012){Abdo}, {Wood}, {DeCesar}, {Gargano},
  {Giordano}, {Ray}, {Parent}, {Harding}, {Miller}, {Wood}, \&
  {Wolff}}]{Abdo2012}
{Abdo}, A.~A., {Wood}, K.~S., {DeCesar}, M.~E., {et~al.} 2012,
\apj, 744, 146

\bibitem[{{Allafort}(2011)}]{AT3207}
{Allafort}, A. 2011, The Astronomer's Telegram, 3207, 1

\bibitem[{{Allafort} \& {D'Ammando}(2010)}]{AT3026}
{Allafort}, A. \& {D'Ammando}, F. 2010, The Astronomer's Telegram,
3026, 1

\bibitem[{{Atwood} {et~al.}(2009){Atwood}, {Abdo}, {Ackermann}, {Althouse},
  {Anderson}, {Axelsson}, {Baldini}, {Ballet}, {Band}, \&
  {Barbiellini}}]{Atwood2009}
{Atwood}, W.~B., {Abdo}, A.~A., {Ackermann}, M., {et~al.} 2009,
\apj, 697, 1071

\bibitem[{{Bonning} {et~al.}(2009){Bonning}, {Bailyn}, {Urry}, {Buxton},
  {Fossati}, {Maraschi}, {Coppi}, {Scalzo}, {Isler}, \& {Kaptur}}]{Bonning2009}
{Bonning}, E.~W., {Bailyn}, C., {Urry}, C.~M., {et~al.} 2009,
\apjl, 697, L81

\bibitem[{{Buehler} {et~al.}(2011){Buehler}, {Scargle}, {Blandford}, {Baldini},
  {Baring}, {Belfiore}, {Charles}, {Chiang}, {D'Ammando}, {Dermer}, {Funk},
  {Grove}, {Harding}, {Hays}, {Kerr}, {Massaro}, {Mazziotta}, {Romani}, {Saz
  Parkinson}, {Tennant}, \& {Weisskopf}}]{Buehler2011}
{Buehler}, R., {Scargle}, J.~D., {Blandford}, R.~D., {et~al.}
2011, ArXiv
  e-prints

\bibitem[{{Buson} \& {Bastieri}(2011)}]{AT3171}
{Buson}, S. \& {Bastieri}, D. 2011, The Astronomer's Telegram,
3171, 1

\bibitem[{{Cannon} \& {D'Ammando}(2010)}]{AT2907}
{Cannon}, A. \& {D'Ammando}, F. 2010, The Astronomer's Telegram,
2907, 1

\bibitem[{{Carini} {et~al.}(1991){Carini}, {Miller}, {Noble}, \&
  {Sadun}}]{Carini1991}
{Carini}, M.~T., {Miller}, H.~R., {Noble}, J.~C., \& {Sadun},
A.~C. 1991, \aj,
  101, 1196

\bibitem[{{Carrasco} {et~al.}(2011){Carrasco}, {Carraminana}, {Escobedo},
  {Recillas}, {Porras}, \& {Mayya}}]{AT3504}
{Carrasco}, L., {Carraminana}, A., {Escobedo}, G., {et~al.} 2011,
The
  Astronomer's Telegram, 3504, 1

\bibitem[{{Ciprini}(2009{\natexlab{a}})}]{AT2049}
{Ciprini}, S. 2009{\natexlab{a}}, The Astronomer's Telegram, 2049,
1

\bibitem[{{Ciprini}(2009{\natexlab{b}})}]{AT2136}
{Ciprini}, S. 2009{\natexlab{b}}, The Astronomer's Telegram, 2136,
1

\bibitem[{{Ciprini}(2009{\natexlab{c}})}]{AT2048}
{Ciprini}, S. 2009{\natexlab{c}}, The Astronomer's Telegram, 2048,
1

\bibitem[{{Ciprini}(2010)}]{AT2943}
{Ciprini}, S. 2010, The Astronomer's Telegram, 2943, 1

\bibitem[{{Ciprini} {et~al.}(2003){Ciprini}, {Tosti}, {Raiteri}, {Villata},
  {Ibrahimov}, {Nucciarelli}, \& {Lanteri}}]{Ciprini2003}
{Ciprini}, S., {Tosti}, G., {Raiteri}, C.~M., {et~al.} 2003, \aap,
400, 487

\bibitem[{{Corbel} \& {Reyes}(2009)}]{AT1933}
{Corbel}, S. \& {Reyes}, L.~C. 2009, The Astronomer's Telegram,
1933, 1

\bibitem[{{Cutini}(2010)}]{AT2669}
{Cutini}, S. 2010, The Astronomer's Telegram, 2669, 1

\bibitem[{{D'Ammando}(2010{\natexlab{a}})}]{AT2860}
{D'Ammando}, F. 2010{\natexlab{a}}, The Astronomer's Telegram,
2860, 1

\bibitem[{{D'Ammando}(2010{\natexlab{b}})}]{AT2783}
{D'Ammando}, F. 2010{\natexlab{b}}, The Astronomer's Telegram,
2783, 1

\bibitem[{{D'Ammando} \& {Vandenbroucke}(2010)}]{AT3002}
{D'Ammando}, F. \& {Vandenbroucke}, J. 2010, The Astronomer's
Telegram, 3002, 1

\bibitem[{{Donato}(2010)}]{AT2583}
{Donato}, D. 2010, The Astronomer's Telegram, 2583, 1

\bibitem[{{Donato} \& {Perkins}(2011)}]{AT3452}
{Donato}, D. \& {Perkins}, J.~S. 2011, The Astronomer's Telegram,
3452, 1

\bibitem[{{Gasparrini}(2011)}]{AT3445}
{Gasparrini}, D. 2011, The Astronomer's Telegram, 3445, 1

\bibitem[{{Gasparrini} \& {Cutini}(2011)}]{AT3579}
{Gasparrini}, D. \& {Cutini}, S. 2011, The Astronomer's Telegram,
3579, 1

\bibitem[{{Ghisellini} \& {Tavecchio}(2008)}]{Ghisellini2008}
{Ghisellini}, G. \& {Tavecchio}, F. 2008, \mnras, 386, L28

\bibitem[{{G{\'o}rski} {et~al.}(2005){G{\'o}rski}, {Hivon}, {Banday},
  {Wandelt}, {Hansen}, {Reinecke}, \& {Bartelmann}}]{Healpix}
{G{\'o}rski}, K.~M., {Hivon}, E., {Banday}, A.~J., {et~al.} 2005,
\apj, 622,
  759

\bibitem[{{Hays} \& {Escande}(2009)}]{AT2316}
{Hays}, E. \& {Escande}, L. 2009, The Astronomer's Telegram, 2316,
1

\bibitem[{{Hays} \& {Marelli}(2009)}]{AT2110}
{Hays}, E. \& {Marelli}, M. 2009, The Astronomer's Telegram, 2110,
1

\bibitem[{{Hill} \& {Vandenbroucke}(2010)}]{AT2413}
{Hill}, A.~B. \& {Vandenbroucke}, J. 2010, The Astronomer's
Telegram, 2413, 1

\bibitem[{{Longo} {et~al.}(2009){Longo}, {Iafrate}, {Hays}, \&
  {Marelli}}]{AT2104}
{Longo}, F., {Iafrate}, G., {Hays}, E., \& {Marelli}, M. 2009, The
Astronomer's
  Telegram, 2104, 1

\bibitem[{{Macomb} {et~al.}(1999){Macomb}, {Gehrels}, \&
  {Shrader}}]{Macomb1999}
{Macomb}, D.~J., {Gehrels}, N., \& {Shrader}, C.~R. 1999, \apj,
513, 652

\bibitem[{{Mariotti}(2011)}]{AT3192}
{Mariotti}, M. 2011, The Astronomer's Telegram, 3192, 1

\bibitem[{{Ojha} {et~al.}(2011){Ojha}, {Dutka}, \& {Torresi}}]{AT3793}
{Ojha}, R., {Dutka}, M., \& {Torresi}, E. 2011, The Astronomer's
Telegram,
  3793, 1

\bibitem[{{Raiteri} {et~al.}(2005){Raiteri}, {Villata}, {Ibrahimov},
  {Larionov}, {Kadler}, {Aller}, {Aller}, {Kovalev}, {Lanteri}, \&
  {Nilsson}}]{Raiteri2005}
{Raiteri}, C.~M., {Villata}, M., {Ibrahimov}, M.~A., {et~al.}
2005, \aap, 438,
  39

\bibitem[{{Rani} {et~al.}(2009){Rani}, {Wiita}, \& {Gupta}}]{Rani2009}
{Rani}, B., {Wiita}, P.~J., \& {Gupta}, A.~C. 2009, \apj, 696,
2170

\bibitem[{{Saz Parkinson} {et~al.}(2010){Saz Parkinson}, {Becker},
  {Carrami{\~n}ana}, {Elsner}, {Harding}, {Johnson}, {Kanbach}, {Odell},
  {Romani}, \& {Swartz}}]{SP2010}
{Saz Parkinson}, P., {Becker}, W., {Carrami{\~n}ana}, A., {et~al.}
2010, in
  Bulletin of the American Astronomical Society, Vol.~42, AAS/High Energy
  Astrophysics Division \#11, 679

\bibitem[{{Schinzel}(2010)}]{AT2829}
{Schinzel}, F.~K. 2010, The Astronomer's Telegram, 2829, 1

\bibitem[{{Schinzel} \& {Ciprini}(2011)}]{AT3670}
{Schinzel}, F.~K. \& {Ciprini}, S. 2011, The Astronomer's
Telegram, 3670, 1

\bibitem[{{Schinzel} {et~al.}(2011){Schinzel}, {Sokolovsky}, {D'Ammando},
  {Burnett}, {Max-Moerbeck}, {Cheung}, {Fegan}, {Casandjian}, {Reyes}, \&
  {Villata}}]{Schinzel2011}
{Schinzel}, F.~K., {Sokolovsky}, K.~V., {D'Ammando}, F., {et~al.}
2011, \aap,
  532, A150

\bibitem[{{Sokolovsky} {et~al.}(2010){Sokolovsky}, {Schinzel}, \&
  {Wallace}}]{AT2402}
{Sokolovsky}, K.~V., {Schinzel}, F.~K., \& {Wallace}, E. 2010, The
Astronomer's
  Telegram, 2402, 1

\bibitem[{{Soldi} {et~al.}(2008){Soldi}, {T{\"u}rler}, {Paltani}, {Aller},
  {Aller}, {Burki}, {Chernyakova}, {L{\"a}hteenm{\"a}ki}, {McHardy}, {Robson},
  {Staubert}, {Tornikoski}, {Walter}, \& {Courvoisier}}]{Soldi2008}
{Soldi}, S., {T{\"u}rler}, M., {Paltani}, S., {et~al.} 2008, \aap,
486, 411

\bibitem[{{Tanaka} {et~al.}(2011){Tanaka}, {Stawarz}, {Thompson}, {D'Ammando},
  {Fegan}, {Lott}, {Wood}, {Cheung}, {Finke}, \& {Buson}}]{Tanaka2011}
{Tanaka}, Y.~T., {Stawarz}, {\L}., {Thompson}, D.~J., {et~al.}
2011, \apj, 733,
  19

\bibitem[{{Tanaka} {et~al.}(2009){Tanaka}, {Takahashi}, \& {Healey}}]{AT2243}
{Tanaka}, Y.~T., {Takahashi}, H., \& {Healey}, S.~E. 2009, The
Astronomer's
  Telegram, 2243, 1

\bibitem[{{The Fermi-LAT Collaboration}(2011)}]{2FGL}
{The Fermi-LAT Collaboration}. 2011, ArXiv e-prints

\bibitem[{{Ulrich} {et~al.}(1997){Ulrich}, {Maraschi}, \& {Urry}}]{Ulrich1997}
{Ulrich}, M.-H., {Maraschi}, L., \& {Urry}, C.~M. 1997, \araa, 35,
445

\bibitem[{{Urry} {et~al.}(1993){Urry}, {Maraschi}, {Edelson}, {Koratkar},
  {Krolik}, {Madejski}, {Pian}, {Pike}, {Reichert}, \& {Treves}}]{Urry1993}
{Urry}, C.~M., {Maraschi}, L., {Edelson}, R., {et~al.} 1993, \apj,
411, 614

\bibitem[{{Wallace}(2010)}]{AT2539}
{Wallace}, E. 2010, The Astronomer's Telegram, 2539, 1

\bibitem[{{Welsh} {et~al.}(2011){Welsh}, {Wheatley}, \& {Neil}}]{Welsh2011}
{Welsh}, B.~Y., {Wheatley}, J.~M., \& {Neil}, J.~D. 2011, \aap,
527, A15

\end{thebibliography}
\end{document}